# Dynamic Road Management in the Era of CAV

Mohamed Younis[1], Sookyoung Lee[2], Wassila Lalouani[1], Dayuan Tan[1] and Sanket Gupte[1],

[1]Dept. of Computer Science and Electrical Engineering, University of Maryland, Baltimore County, Baltimore, MD, USA
Email: younis, lwassil1, dayuan1, sgupte1@umbc.edu

[2]Dept. of Computer Science and Engineering, Ewha Womans University, Seoul, Korea,
Email: sookyounglee@ewha.ac.kr (corresponding author)

### Abstract

Traffic management and on-road safety have been a concern for the transportation authorities and the engineering communities for many years. Most of the implemented technologies for intelligent highways focus on safety measures and increased driver awareness, and expect a centralized management for the vehicular traffic flow. Leveraging recent advances in wireless communication, researchers have proposed solutions based on vehicle-to-vehicle (V2V) and vehicle-to-Infrastructure (V2I) communication in order to detect traffic jams and better disseminate data from on-road and on-vehicle sensors. Moreover, the development of connected autonomous vehicles (CAV) have motivated a paradigm shift in how traffic will be managed. Overall, these major technological advances have motivated the notion of dynamic traffic management (DTM), where smart road reconfiguration capabilities, e.g., dynamic lane reversal, adaptive traffic light timing, etc. will be exploited in real-time to improve traffic flow and adapt to unexpected incidents. This chapter discusses what the challenges in realizing DTM are and covers how CAV has revolutionized traffic management. Moreover, we highlight the issues for handling human-driven vehicles while roads are transitioning to CAV only traffic. Particularly, we articulate a new vision for inter-vehicle communication and assessment of road conditions, and promote a novel system for traffic management. Vehicle to on-road sensors as well as inter-vehicle connectivity will be enabled through the use of hand-held devices such as smartphones. This not only enables real-time data sharing but also expedites the adoption of DTM without awaiting the dominant presence of autonomous vehicle on the road. The proposed traffic management system incorporates computation and communication capabilities in traffic lights and road-side units, and accounts for the human-factor in controlling the traffic flow. The goal is to allow autonomous and driver-centric routing decisions that not only are locally optimal but also serves an overall system objective. Sample results of some of our on-going work are also presented. Open research issues are further outlined.

## 1 Introduction

### *1.1 Road Traffic Problems*

Vehicular traffic congestion has become a daily problem that most people suffer from, especially in urban areas. With the increasing number of vehicles on the roads, slow traffic and congestion have become an unpleasant expectation during daily commutes. According to Nationwide insurance company, "The average urban commuter is stuck in traffic for 34 hours every year and 1.9 billion gallons of fuel, more than five days' worth of the total daily fuel consumption in the United States were wasted due to road congestion" [1]. Worldwide, the average commuter spent an extra 100 hours a year travelling during the evening rush hour alone in 2014 and the number hits 272 hours in 2018 in the Columbia capital, Bogota, where commuters experience the world's greatest traffic jams [2]. Moreover, in the largest urban areas across the United State, commuters consume nearly 7 full working days in extra traffic delay in 2017, which is equivalent to over $1,000 in personal costs [4]. This not only impacts productivity but also poses a safety hazard. In addition, traffic congestion causes excessive fuel consumption and high doses of pollution, which

adds to the negative economic impact on the nation and potential health risks for citizens. According to the 2019 Urban Mobility Report of the Texas A&M Transportation Institute [2], congestion has been persistently growing and is not restricted to large metropolitan areas as indicated by the statistics in Figure 1 and the average auto commuter spends 54 hours in congestion and wastes 21 gallons for fuel due to traffic congestion, which translate to $1,010 of congestion cost per auto commuter. In 2017, the overall congestion cost in urban areas is about $166 billion due to the extra 8.8 billion hours trip which requires purchase of an extra 3.3 billion gallons of fuel.

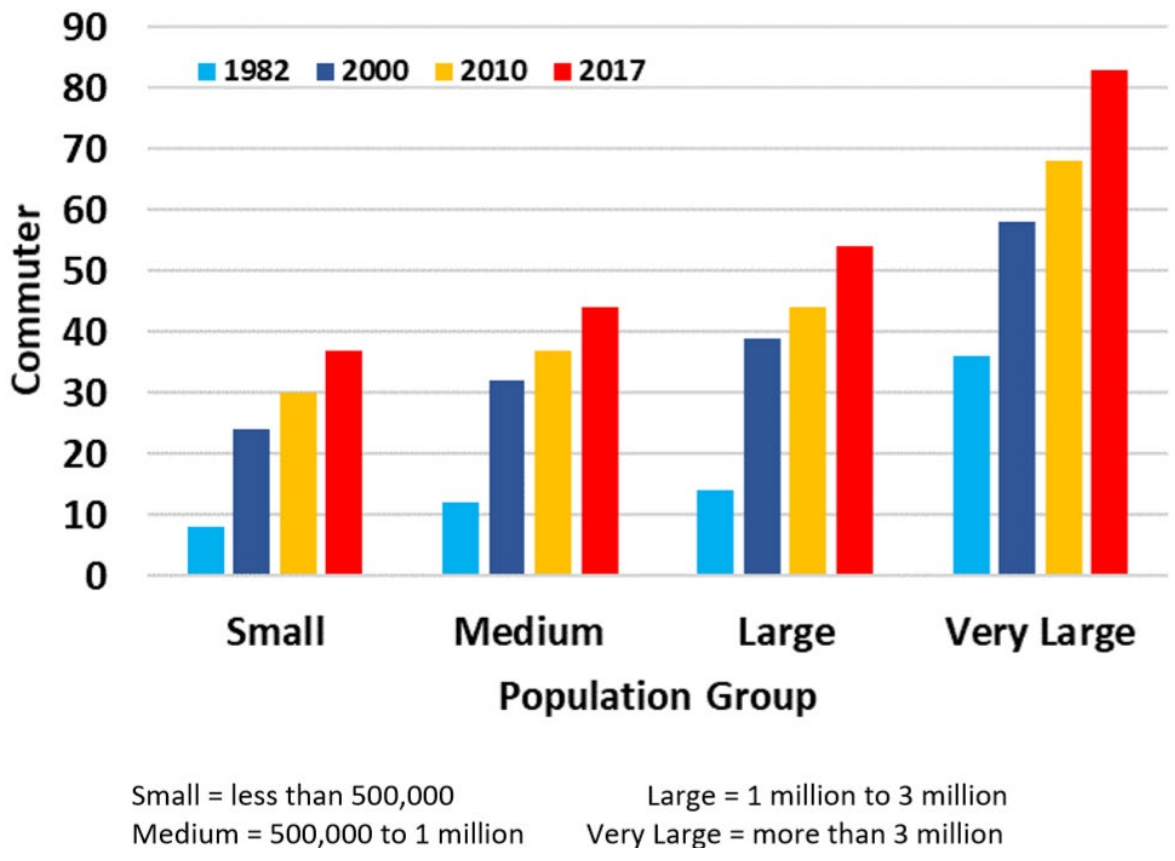

**Figure 1**: With all measures to alleviate traffic congestion, little progress has been made; the effect has stabilized rather than reduced. Plot is from [4].

### *1.2 Conventional and Emerging Congestion Mitigation Methodologies*

The transportation community has come to realize that balancing the traffic load on existing roads is a key for effective mitigation of congestion [5]. However, the steps taken by authorities either are constrained by driver's response, e.g., by fostering ride sharing and encouraging the use of mass transient [6]-[9], or employ measures that do not adapt based on real-time conditions, e.g., impose a road use pattern following a static schedule that is based on time of day and day of the week. For example, many cities throughout the U.S. designate High-Occupancy Vehicles (HOV) lanes, and charge tolls to cross tunnels and bridges, and to drive over certain segments of highways. Yet, these measures prove ineffective since the driver choice to avoid expensive routes is more or less unpredictable, as it is difficult to make an educated decision on trading off time and cost while being behind the wheel. Other cities such as Washington DC, change the direction of lanes on certain roads during rush hours, while others (e.g., Minneapolis and Seattle) control the vehicle entry rate to highways through ramp metering during these hours. Again, this static schedule factors in road capacity as the main cause of congestion and does not factor in all incidents, traffic light timing, weather, the effect of these measures on local roads, etc. In addition, conventional means for alerting drivers (e.g., signs and radio updates) lack responsiveness and do not mitigate or prevent traffic congestion.

Given the shortcoming of static schemes, the notion of active (dynamic) traffic management has gained lots of attention in recent years. The key features of the dynamic traffic management (DTM) paradigm are exploiting interaction with the drivers (or vehicles) to predict traffic jams and proactively employing means to avoid them. Examples of DTM based congestion mitigation schemes include [10]:

- *Adjustable shoulder use:* converting road shoulder into a lane in response to congestion or accident in order to increase throughput.
- *Varying speed limits:* setting the speed limit based on the road condition and vehicle density.
- *Adaptive ramp metering:* adjusting the timing of traffic signals at ramp entrances to control vehicle in-flow to highways.
- *Dynamic rerouting:* directing traffic to route alternatives in order to prevent congestion.
- *Adaptive traffic signal timing:* varying the traffic light timing and/or phases to improve throughput and delay at an intersection.

A shared characteristic among these unconventional DTM based schemes is that they involve some form of road reconfiguration, which is a revolutionary view of such major infrastructure.

### *1.3 Connected Vehicles and Infrastructure*

In the early transportation system, a wide variety of traffic monitoring technologies using sensing technologies such as safety CCTV, traffic video cameras, piezo-electric sensors, inductive loops have been introduced to monitor road conditions and alert motorists through electronic variable-message signs. However, due to lack of sufficient coverage and high maintenance cost, the transportation systems have evolved by using various types of wireless and mobile technologies such as 2G/3G/4G/LTE/5G, Wi-Fi, GPS, etc. for real-time traffic monitoring [11][12][13]. Then, intelligent transportation systems (ITS) have emerged with the advances of information and communication technology, and the prospect of leveraging recent developments in the vehicular ad hoc network (VANET) and wireless sensor network (WSN) areas. Furthermore, it is highly expected in academia and automotive industry that fully autonomous vehicles (AV) will be fulfilled between 2025 and 2030 and their global dispersion become fact between 2030 and 2040 [14][15]. The impact of AVs includes reduced traffic, infrastructure saving such as parking congestion, increased safety, energy conservation and pollution reductions, and independent mobility for low-income people [16].

Moreover, the remarkable research results coming from the fields of in-vehicle digital technology, wireless communication, embedded systems, intelligent routing system, sensors and ad-hoc technologies have given rise to the emergence and evolution of connected autonomous vehicles (CAV). The advent of CAV will lead to a paradigm shift of automobile design from an old-fashioned source of repositioning into a full-scale, smart, and infotainment-rich computing and commuting device. In contrast to human-driven vehicles, CAVs cooperatively share the road that they travel on and can thus be controlled to adaptively handle increased vehicle density and be provided with routes to dynamically optimize the delay for the individual travellers and the vehicular throughput on the road network. Consequently the arrival of CAV will change the model for how road traffic will be managed and how congestion could be mitigated, and will eventually provide travellers with more safe, accurate, timely decision during a road trip reducing human errors and life-threatening situation on the road [17][18]. In other words, CAV will enable the full realization of the DTM concept.

### *1.4 Scope and Organization*

This chapter introduces the reader to the notion of dynamic traffic management in the context of ITS. Particularly, the complications in realizing the full potential of dynamic traffic management are discussed and how CAV can be instrumental in overcoming these complications. We highlight the various wireless communication technologies for supporting vehicle to vehicle (V2V) and vehicle to infrastructure (V2I) interaction. We then analyze and enumerate the attributes that can be shared through CAV with the road management infrastructure in order to enable dynamic adaptation of the road configuration as a means for optimizing traffic flow and improving both traveler-centric and system-based performance metrics. Existing CAV-enabled adaptive road reconfiguration techniques are categorized into five groups, namely, autonomous intersection management (AIM), adaptive traffic light control (ATLC), dynamic lane grouping (DLG), dynamic lane reversal (DLR) and dynamic trajectory planning (DTP). We describe each category in detail and compare the various techniques. We further highlight the issues when autonomous and driver-based vehicles co-exist on the road and the impact of such a mix on the various road management strategies. Finally, we present our vision for how vehicular traffic will be managed in smart cities and discuss our Internet of Radio-equipped On-road and vehicles-carried Agile Devices (iRoad) project for realizing such a vision. We also report on the results of some of our on-going research and outline future research topics that warrant more investigation.

The chapter is organized as follows. The next section highlights the challenges in implementing dynamic traffic management. Section 3 focuses on how CAV can facilitate DTM, and categorizes existing techniques in that regard. In Section 4, we highlight the challenge of incorporating DTM in the presence of human-driven vehicles on the road and discuss efforts within our iRoad project for overcoming these challenges. Finally, Section 5 concludes the chapter and outlines open research problems.

## 2 DTM Challenges

As pointed out earlier, dynamic traffic management opts to respond to incidents and/or congestion in real-time. In essence, DTM models a road network as a closed-loop cyber-physical system and ideally provisions the means for autonomous control of such a system. In this section, we highlight the challenges in realizing DTM in practice. The main issues are discussed in the following subsections.

### *2.1 Data Collection*

To respond to road incidents and congestion, the status of traffic has to be continually tracked. Traffic condition assessment methodologies can be categorized based on the accuracy of the data collection and on what the road status is needed for. From the vehicle (driver or passenger) perspective the data is used to detect congestion, estimate arrival time, and decide on the best travel route. On the other hand, a local branch of the department of transportation will be interested in using the data for predicting traffic jams and deadlocks, performing analytics to measure utilization and assess criticality of road infrastructure, providing alerts, and diverting vehicles to alternate routes if needed. There are quite a few traffic monitoring systems that gather and disseminate traffic information. These systems can be classified as: (i) infrastructure based that are installed and controlled by the authority, and (ii) participatory where the data is voluntarily provided by participants or indirectly inferred from other context. Example of infrastructure bases monitoring systems are on-road sensors, e.g., traffic cameras, loop detectors, laser sensors, and pressure hose. Traffic cameras are the most popular on-road sensors where not only an administrator can get a visual view of the conditions but also computer vision techniques can be applied to recognize and count vehicles in the live video [19][20]. Electromagnetic loops and laser sensors are popular at intersections and are used to determine the traffic signal sequence [21]. Radar is also used at intersections for not only detecting vehicles but also counting them so that the signal timing and sequence are optimally adjusted [22]. Pressure hoses [23] are typically used during field studies to count vehicles and estimate traffic intensity on a road segment during certain duration; generally they are not durable and not intended for real-time monitoring. Overall, infrastructure based monitoring systems are expensive, mainly because of the installation cost.

Participatory systems, on the other hand, either: (i) exploit the popularity and recent advances in wireless technologies, (ii) collect location and contextual data that is voluntarily provided by drivers, or (iii) leverage the wealth of vehicle's onboard sensors [24]. For example, Zhang et al. [25] utilize wireless mesh networks to track the movement of specific vehicles; these vehicles are roaming the roads and responding to wireless probes. By localizing the probe responses using mesh relays the vehicles can be located and their motion pattern and delay can then be correlated to estimate the conditions of the travelled roads. Similarly, routinely traveling vehicles such as buses and taxis are utilized in [26] to report on-road traffic. Prime examples of traffic monitoring systems that rely on voluntarily provided data are Google maps, Waze [27], Inrix [28], and Cellint [29], where the location of mobile individuals is determined through the GPS on their cell phones or portable computing devices while riding their vehicles.

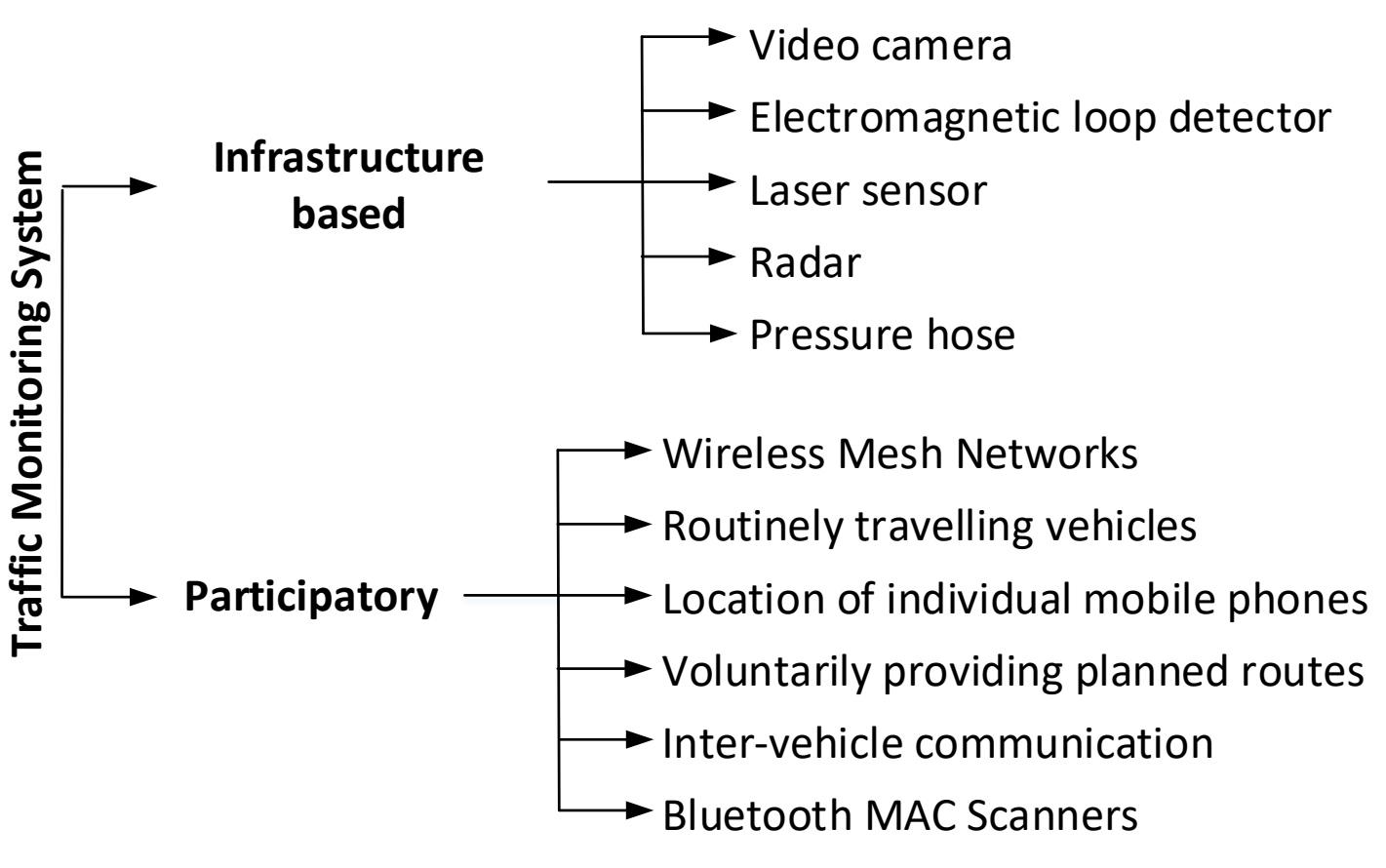

**Figure 2**: Categorization of the data collection methodologies and underlying means.

By correlating the location information, the motion speed and vehicle density are estimated to infer traffic conditions on road segments [30], and queue lengths are calculated at intersections [31]. Some work even assumes that a traveler will voluntarily provide the primary route and alternatives, which enables predicting vehicle density over time [32]. Meanwhile, systems like TrafficView [33][34] and SOTIS [35] rely on inter-vehicle communication in collecting data. Some work has focused on just detecting congestion [36]-[50]. The detection methodologies vary from a simple monitoring of motion speed [36], to conducting simulation using mobility traces [42][43] or applying fuzzy logic and generic algorithms [37][44]-[50]. Some approaches try to devise a congestion prediction model as well [51]-[53]. Finally, multiple modality has been exploited to improve the fidelity of the traffic assessment. For example, Bluetooth MAC Scanners (BMS) is exploited in [54] as an extra modality to boost the accuracy and reliability of the traffic flow measurements made by loop detectors. Other systems, e.g., [38], assume the availability of a high-level traffic report and use the vehicle's local observations for fine-grained assessment, e.g., by checking travel speed of other vehicles. Figure 2 provides a summary of the popular means for collecting traffic data.

### *2.2 Road Configuration*

The notion of road parameters reconfiguration is analogous to adjusting the capacity of and controlling flow over links in communication networks. By modeling the road infrastructure as a network, one may apply the well-established graph theoretical algorithms to study the performance and predict problems. Particularly, applying network flow analysis techniques will enable estimating throughput, identifying bottlenecks and determining best means for stabilizing the operation and maximizing performance. Road configurability is realized in practice by means such as (i) traffic signal timing, (ii) ramp metering, (iii) tolls, (iv) speed limit, (v) HOV lanes, (vi) shoulder, and (v) contraflow lanes. The first four, namely, signal timing, ramp metering, tolls and speed limit, mainly control the flow to reduce delay, and are exploited to prevent congestion in certain travel direction. For example, ramp metering is used to control the in-flow to a highway from local roads in order to mitigate slow down when the vehicles merge after entering the highway and also avoid exceeding the highway capacity. On the other hand, shoulder use and lane reversal boost the capacity of the road, and in case of lane reversal, the increase will be at the expense of reduced capacity in the opposite direction; the goal is to increase vehicle throughput, and consequently passenger throughput. Meanwhile, the designation of HOV lanes opts to improve passenger throughput only.

Road configuration parameters are currently set based on time of day and day of the week. Typically, statistics for traffic intensity are used to determine the expected conditions and consequently what values are to be assigned to the various parameters in order to optimize contemporary metrics like vehicular throughput and delay. Safety is also factored in, particularly when it comes to intersection crossing and speed limit settings. The statistics are based on historical data collected during normal circumstances, i.e., in the absence of traffic incidents such as accidents. To realize DTM, road reconfiguration is to be exploited autonomously and in real-time based on the traffic status and trend. In other words, the road parameters have to be adaptively adjusted to cope with variations in the traffic patterns. Such an approach will enable effective handling of emerging events that are often experienced sporadically with no predictable patterns. To elaborate, collisions and vehicle breakdown incidents often create traffic jams and may happen at any time. Being able to ease the impact of traffic incidents will be invaluable for both motorists and authorities.

There have been some efforts for supporting optimized road reconfiguration, yet with limited scope. Some approaches exploit dynamic pricing to divert traffic from certain road segments by announcing a toll hike [55]-[59] and making the speed limit variable to improve flow [60][61]. However, the response is typically slow since the adjustment is centrally controlled and determined by the authority. Some consider road configurability at the planning stage by determining whether HOV lanes should be employed [62]. Adaptive traffic light scheduling is also pursued to deal with vehicle pileup at individual intersections [63]-[68]. However, the approach does not factor in the impact on other parts of the road. Very few studies, e.g., [69]-[72], have explored coordination among traffic lights to increase traffic flow; however, none of them considers possible road reconfiguration by changing lane direction. VANET has also been exploited as a means to orchestrate intersection crossing for self-driving cars [73]-[75]. Finally, the focus of [76] is limited to traffic signal timing. A comprehensive DTM optimization model that factors all means for road reconfiguration is yet to be developed.

### *2.3 Communication and Control*

DTM can be implemented in a centralized or a distributed manner. As pointed out above, there is no optimization models that factor all means for road reconfiguration; we further note that centralized control has conventionally been assumed by existing work on DTM. To enable distributed DTM implementation as well as support on-road data collection, means for V2I communication has to be provisioned. Given the vehicle mobility and also to avoid the prohibitive cost of wiring, wireless technologies are considered the default for establishing communication links. Basically, connectivity is needed to support interaction among vehicles and between them and road-based data collection and configuration controllers. Communications among the various road units could be wired or wireless; yet wireless links are way less costly to provision for. Interaction among road units could be for coordinated control of the road configuration and for sharing data. In the following we enumerate popular wireless technologies and analyze their applicability in DTM systems. Table 1 summaries and compares their features.

- *Long Range WiFi:* This technology extends the range of the popular WiFi which does not exceed 100 meters in outdoor setups. By employing directional antennas, Long Range WiFi achieves a range of multiple kilometers [77]. Other advantages of Long Range WiFi include the use of unlicensed spectrum, the incorporation of small and inexpensive antennas, and the availability of reliable and free-licensed software, e.g., DD-WRT [78]. Long Range WiFi is suitable for communication among road configuration units, e.g., between controllers of consecutive traffic signals in order to coordinate timing.
- *Cellular Telecommunication*: For a communication range in excess of 10 kilometers, Long Range WiFi is not a viable option. In this case cellular network is a more appropriate choice that enables the establishment of reliable connections and is supported by well-established service providers. The radius of a cell varies from 1 to 30 kilometers. Yet, the reliance on base-stations could constitute an obstacle in low coverage areas and introduce high latency during heavy network loads [79].
- *IEEE 802.11p*: This IEEE standard is mainly developed to support V2V and V2I communication. The vision is that it serves as the wireless backbone for ITS [80]. The range of the IEEE 802.11p is capped to 1 kilometer [81]. It also is able to support data exchange among fast moving vehicles.
- *Dedicated Short Range Communication (DSRC)*: DSRC is based on IEEE 802.11p, is designed to boost on-road safety through the exchange of messages among vehicles. Several alerts are shared among vehicles to avoid collisions, such as Forward Collision Warning (FCW), Emergency Electronic Brake Lights (EEBL), Blind Spot Warning (BSW), Do Not Pass Warning (DNPW), Intersection Collision Warning (ICW), etc. [82]. DSCR also provides a Basic Safety Messaging (BSM) mechanism, which broadcasts status information of each vehicle, including position, speed, acceleration and direction, at a frequency of 10 times every second over a range of a few hundred meters [82]. The safety message data could be leveraged by the road units as well. For example, a traffic signal controller could be augmented with a DSRC transceiver to overhear safety messages in the vicinity and assess the in-flow vehicle volume in the various direction and the out-flow rate. Such assessment can then be used to dynamically set the green time in order to maximize throughput and reduce vehicle waiting time at an intersection.
- *WiFi Direct*: This technology, which is also referred to as WiFi Peer-to-Peer, is used for near field communication to support data exchange within a range of a few hundred meters. It offers the data rate of a typical WiFi and its support is becoming standard nowadays on smart devices [83]. The distinct feature of WiFi direction is that it does not need a wireless router and enables devices to establish peer-to-peer links and dynamically form groups. Like DSRC, WiFi direct can be used to support communication among vehicles and with close-by road units, especially when traffic involves conventional vehicles and the driver's cell phone is used to establish communication links [84].

- *Bluetooth*: Like WiFi Direct, Bluetooth is geared for device-to-device communication. Despite the popularity of Bluetooth, it suffers limitations that diminish its suitability for the realization of DTM systems. Basically, the communication range of Bluetooth is less than around 100 meters and supports at most eight connections. It can serve as a secondary means for communication between closely located vehicles.

## *2.4 Traffic Assignment*

Traffic assignment refers to how vehicles are routed and plays a profound role in forecasting travel time. In essence, it is a means for controlling the vehicle density to manage traffic and optimize performance. In other words, traffic assignment constitutes the action for closing the traffic control loop. Traffic assignment opts to optimally allocate a set of origin-destination (O-D) pairs to a specific set of paths, i.e., consecutive road segments, according to criteria set by the system and drivers. The optimization objective could be minimizing the travel distance, maximizing the vehicular throughput, or minimizing fuel consumption. The considered constraints include the infrastructure capacity, safety rules, and traffic regulations. Traffic assignment is generally a very complex optimization since it involves allocating road resources, e.g., lanes, and scheduling vehicles entry. Such optimization problem is NP-hard if done statically, let alone the complexity when conducted in real-time where the vehicle arrival rate fluctuates and traffic incidents sporadically take place with no regular pattern. In other words, traffic assignment in DTM has to be formulated as a time-dependent optimization problem.

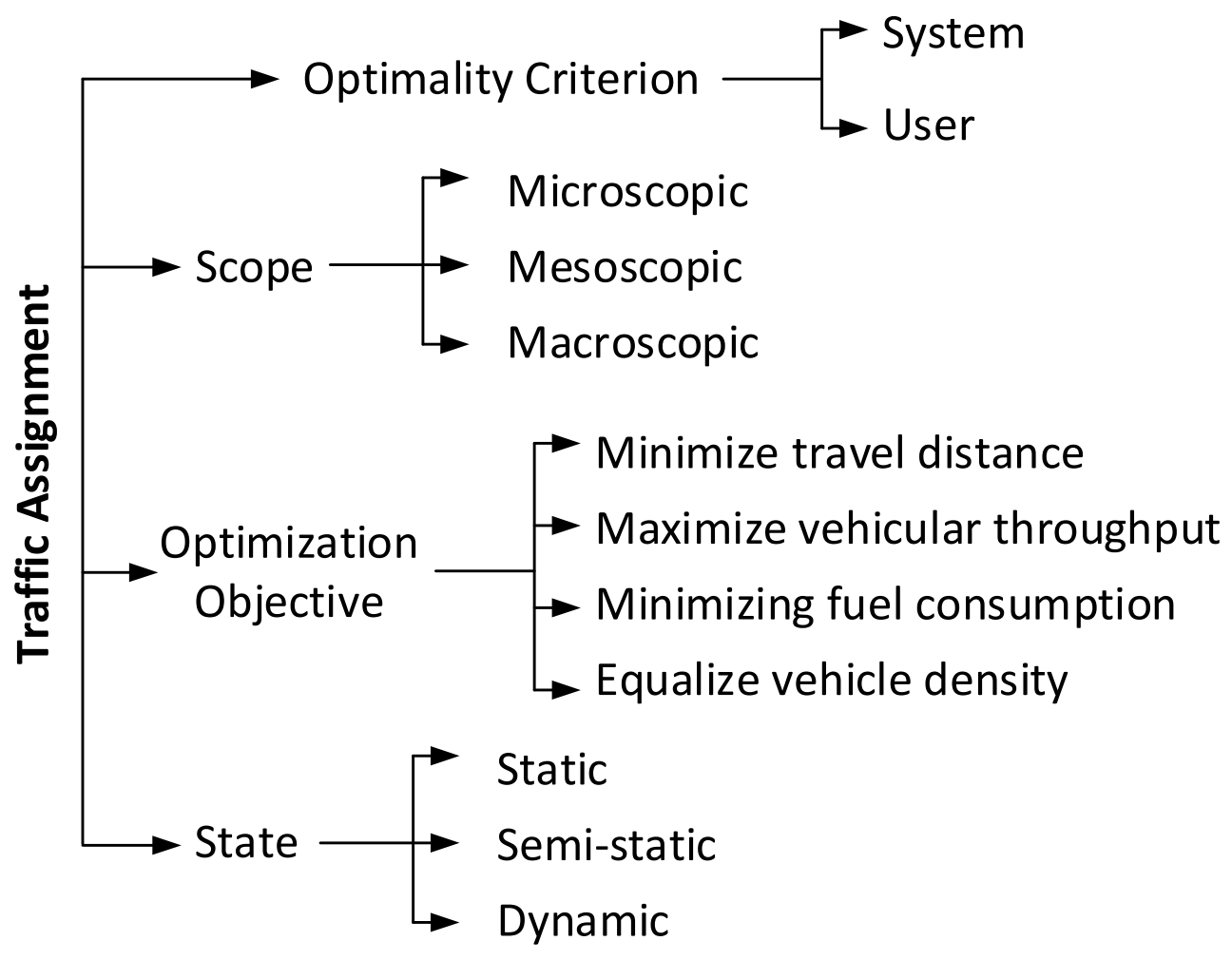

**Figure 3:** General Taxonomy of the categorization of the various traffic assignment techniques.

Figure 3 highlights the various classifications of the traffic assignment problem. As indicated in the figure, the traffic assignment optimization can be geared for system level optimality criteria where the big picture matters the most. For example, road capacity utilization could be the main worry, even at the expense of causing inconvenience to some road users. Achieving such optimization objective requires means for controlling traffic flow and vehicle density either through on-road signs/signals, e.g., regulating in-flow rate, changing lane designation, etc., or influencing user selection, e.g., by varying the toll charges. Meanwhile, the objective could be user centric where the road experience of individuals is targeted. For example, the least arrival time may be the quest of a user regardless whether the picked route serves a global optimization metric or not. User centric strategies are the most popular in case of traffic involving human-driven vehicles. When vehicles collectively try to improve the experience of all users, the objective is called dynamic system optimum [85]. The latter is a perfect match for CAV based scenarios.

**Table 1:** Comparison of Different Communication Technologies.

| Communication Technologies | Range | Frequency | Responding Speed | Need/ Support Relay or not | Support Multiple Vehicles | Applications | Pros | Limits |
|---|---|---|---|---|---|---|---|---|
| **Long Range WiFi** | Max 315 km in practise use. | 2.4 GHz or 5.8 GHz | Up to 356.33 Mbit/s | Support | Yes. Support MIMO. | 1. Bring Internet to remote construction sites or research labs.<br>2. Connect widespread physical guard posts, e.g. in forest.<br>3. Transmit real time seismic data in Peru to UCLA. | 1. Long range.<br>2. Support MIMO.<br>3. Unlicensed spectrum.<br>4. Smaller, simpler, cheaper antennas (compared to cell or fixed antennas). | 1. Line of Sight limitation.<br>2. Limited to soft obstacles.<br>3. Lack of commercial service providers. |
| **Cellular Network (LTE)** | For optimal performance is <5 km.<br>For reasonable performance is <30 km.<br>For acceptable performance is <100 km. | 450 MHz - 3700 MHz | Downlink peak rates of 300 Mbit/s, uplink peak rates of 75 Mbit/s.<br><br>Latency is <70 ms, up to <5 ms. | Need | Yes. Support MIMO. | Cellular Netowrk | 1. Large range.<br>2. Support MIMO.<br>3. Support for terminals moving at up to 500 km/h (310mph).<br>4. Support for cell sizes from tens of metres radius (femto and picocells) up to 100 km (62 miles) radius macrocells.<br>5. Support of at least 200 active data clients in every 5 MHz cell. | Needs base stations. |
| **Cellular Network (5G)** | A few hundred meters. | Below 6 GHz, and 24 GHz - 300 GHz | 2Gbit/s.<br>Latency is <30 ms, up to <1 ms. | Need | Yes. Support massive MIMO. | Celluar Netowrk. Internet of Things. | 1. High data rates.<br>2. Low latency.<br>3. Energy saving.<br>4. Large system capacity and large-scale device connectivity.<br>5. Support massive MIMO. | 1. Need high density base stations.<br>2. Short range.<br>3. Poor ability to pass through building walls. |
| **IEEE 802.11p** | 1 km.<br>Support moving vehicles up to 260 km/h (163 mi/h). | 5.85 - 5.925 GHz | Quicker than legacy WiFi.<br>No need to wait on the association and authentication procedures to complete prior to exchanging data. | No | Yes | Toll collection, vehicle safety services, and commerce transactions via cars. | 1. Support data exchange between high-speed vehicles and between the vehicles and the roadside infrastructure.<br>2. Created specially for Intelligent Transportation Systems (ITS)). | No authentication and data confidentiality mechanism. |
| **Dedicated short-range communications (DSRC)** | <450 km | 5.9 GHz | Max 150 ms latency | No | Yes | Used in electronic toll collection (ETC) in Europe and Japan. | 1. Have multiple channels:<br>one control channel (CCH),<br>six service channels (SCHs).<br>2. Support vehicles moving in high speed. | Short range. |
| **WiFi Direct** | <200 m | 2.4 GHz, 5 GHz | <250 Mbps | No | Yes | 1. large files transfer<br>2. Connection between printers, cameras, scanners, wireless mice and many other common devices. | 1. No need require and confirm.<br>2. No need wireless router (AP).<br>3. Support diff manufactures.<br>4. High speed (than bluetooth). | Short range. |
| **Bluetooth** | 10 - 300 m | 2.4 - 2.8 GHz | 6 - 100 ms | No | Support max 8 connections. | Connection between printers, scanners microphones and so on. | Send small snippets of information even not paired or connected. | Short range. Slow. |

The granularity of DTM depends on the underlying traffic flow model. Generally, traffic flow models can be classified into three categories: macroscopic mesoscopic, and microscopic [86][87]. As the name indicates, microscopic models are fine-grained and factors in vehicle-level behavior, and vehicle-to-vehicle and vehicle-to-road interactions [88]. In case of human-driven vehicles the driver's response to incidents and traffic conditions is captured as well. For example, traffic flow in a construction zone is significantly influenced by drivers, e.g., tailgating, passing speed, etc. Generally microscopic models involve excessive details and, if adopted, would complicate the traffic assignment process. They could be more suited for CAV given the autonomous control of the involved vehicles. Macroscopic models, on the other hand, are more coarse-grained and categorize traffic in an aggregate term, e.g., average motion speed, and average vehicle density [89]. Aghamohammadi and Laval [90] have further classified macroscopic models as continuous and discrete space. The former abstracts the traffic assignment problem to operate on regions while the latter models the area as road segments (finite number of zones). Continuous-space models are useful when the network is dense such that both the distance between road intersections and their longitude is small compared to the size of the region. Macroscopic models are more popular for managing traffic on highways, major local roads, city-based street grids, etc. [91]. A mesoscopic model falls in between the microscopic and macroscopic ones where the individual vehicles are considered yet the traffic attributes are captured through a probability distribution function [92].

Whether the traffic assignment is based on user equilibrium or system optimality, popular objectives of the optimization include minimizing the travel time, minimizing the driving distance (shortest path to destination), minimizing fuel consumption, maximizing vehicular throughput, and equalizing the traffic density on the roads [93][94]. The last two objectives are more common for system level optimization than for individual vehicles. Multiple objectives could also be pursued where a weighting function is employed to reflect the level of importance. Usually the road network is modeled as a graph with link cost that reflects the optimized attributes. For example, when the travel time is the target of optimization, the average delay for travelling on a road segment will be used as the cost associated with the corresponding link in the graph. The basis for assessing the link cost may be deterministic or stochastic. Deterministic means relies on real-time traffic data collected by on-road sensors, e.g. traffic cameras, or through V2V communications. Stochastic means includes probability distributions, i.e., mesoscopic models, or historical data sets. Moreover, DTM is an iterative process for a defined traffic flow and specific period of time. Thus, the link cost is a function of time. The complexity of solving the traffic assignment optimization varies widely based on the objective and link cost functions. For deterministic (scaler) link costs and a linear objective function, a user could apply classical least-cost routing algorithms such as Dijkstra's and Bellman-Ford. When system-level metrics are targeted, the problem is often mapped to multi-commodity flow optimization. Such a problem is generally NP-hard; some variants could have polynomial time solutions [95].

Traffic assignment models are classified based on the temporal dimension into three categories: static, semi static and dynamic. The difference among them is based on whether the modeling of the flow considers congestion and captures whether there are variations between the in and out flow for each road segment or zone within the area. The dynamic category reflects instantaneous real-time reaction to incidents and peak rates of vehicle entry to the road network. Therefore, traffic data should be accurate and fresh in order for the complexity of the dynamic strategy to be justified. Dynamic strategies are well-suited for CAV. The semi-static category differs from the dynamic one in the frequency at which situations are assessed and actions are taken; in essence it strikes a balance between responsiveness and complexity.

## 3 CAV-enabled Traffic Management

Dynamic traffic management at intersections have been the main focus in urban areas since junctions are often bottlenecks in road networks. DTM strategies in that context generally follow two methodologies, namely, time management and space management [96]. Work on time management at intersections can be categorized into two groups: (a) traffic light phasing and timing control at signalized intersections, and (b) autonomous intersection management for controlling semi- or fully autonomous vehicles at signalized or unsignalized intersections. Meanwhile, space management is generally done through road reconfiguration and is deemed to be instrumental for improving vehicular traffic throughput. The idea is to increase road utilization by (i) dynamic lane grouping which adaptively reassigns turn movements to lanes depending on real-time traffic demands, (ii) adaptively reversing contraflow lanes by considering changes in the traffic flow volume, and (iii) dynamic trajectory planning to factor in coordinated vehicle motion. In the balance of this section we discuss these techniques and summarize the state of the art.

### *3.1 Autonomous Intersection Management (AIM)*

Intersection management strives to optimize cycle time, splits, and offsets of traffic light signals. In the case of CAVs, vehicle's arrival and request for green light at intersections can be approximately predicted along with its routes. Such prediction is possible since the vehicle may share real-time locations and could be guided by an infrastructure-based controller. Therefore, the problem of the intersection management in CAV is significantly different from the traditional methods. The main focus of intersection management for self-driving vehicles is on eliminating the potential overlaps of vehicles coming from all conflicting lanes at an intersection and improving passengers' safety and fairness as well as stopping delay, fuel consumption, air quality and total travel time in comparison to the conventional actuated intersection control using traffic signals and stop signs. In order to achieve the objectives, inter-vehicle cooperation and/or V2I communication are required for effective intersection operations and management. Controlling and managing CAVs at an intersection can be conducted in a centralized manner, where a single central controller globally decides for all vehicles [126][127], or decentralized where each vehicle determines its own control policy based on the information received from other vehicles on the road, or from a coordinator using V2V or V2I communication [128][129][130]. For example, the distributed auction-based intersection management approach of [128] employs an automatic bidding system that operates on behalf of the driver. The bidding system is applied at traditional intersections, i.e., stop signs and traffic signals, based on trip characteristics, driver-specified budget, and remaining distance to the destination.

In reality, achieving vehicle's full autonomy is not expected to be instantaneous and vehicles have been gradually equipped with more and more advanced driver assistance systems (ADAS). Therefore, considering semi-autonomous vehicles raises issues for AIM systems. In particular, under heavy traffic and difficult driving situations, current ADAS technologies are unable to handle certain dangerous scenario either at intersections or when merging on highways, most notably when dealing with vehicles arriving from the sides. Therefore, a cooperative framework has been proposed for semi-autonomous vehicles to mitigate the risk of collision or deadlocks while remaining compatible with conventional scenarios involving human-driven vehicles [131]. Another semi-autonomous intersection management allows vehicles with features such as adaptive cruise control to enter an intersection from different directions simultaneously and achieves great reduction in traffic delay at an intersection [132].

In addition, early AIM work has focused on protecting passengers by seeking a safe maneuver for every vehicle approaching an intersection without considering traffic lights, and mitigating possible system failure cases that could result from inevitable trajectory overlaps at the intersection [133]-[135]. Another trajectory-based AIM system optimizes traffic light signal control simultaneously with the autonomous vehicle trajectories based on real-time collected arrival data at detection ranges around the center of the intersection [136]. Deadlocks and starvation (unfairness) are concerns that have also been tackled, where lightweight optimization of trajectories for safe and efficient intersection crossing are proposed [129][137]. Some of the AIM systems focus on minimizing fuel consumption subject to throughput and safety requirements. The throughput maximization problem with hard safety constraints has been formulated as a decentralized optimal control problem for fuel minimization; an analytical solution has been presented such that vehicles pass an intersection without a full stop and each vehicle's acceleration and deceleration are optimized [130]. Consequently, transient engine operation is minimized and fuel consumption is saved while also improving travel time.

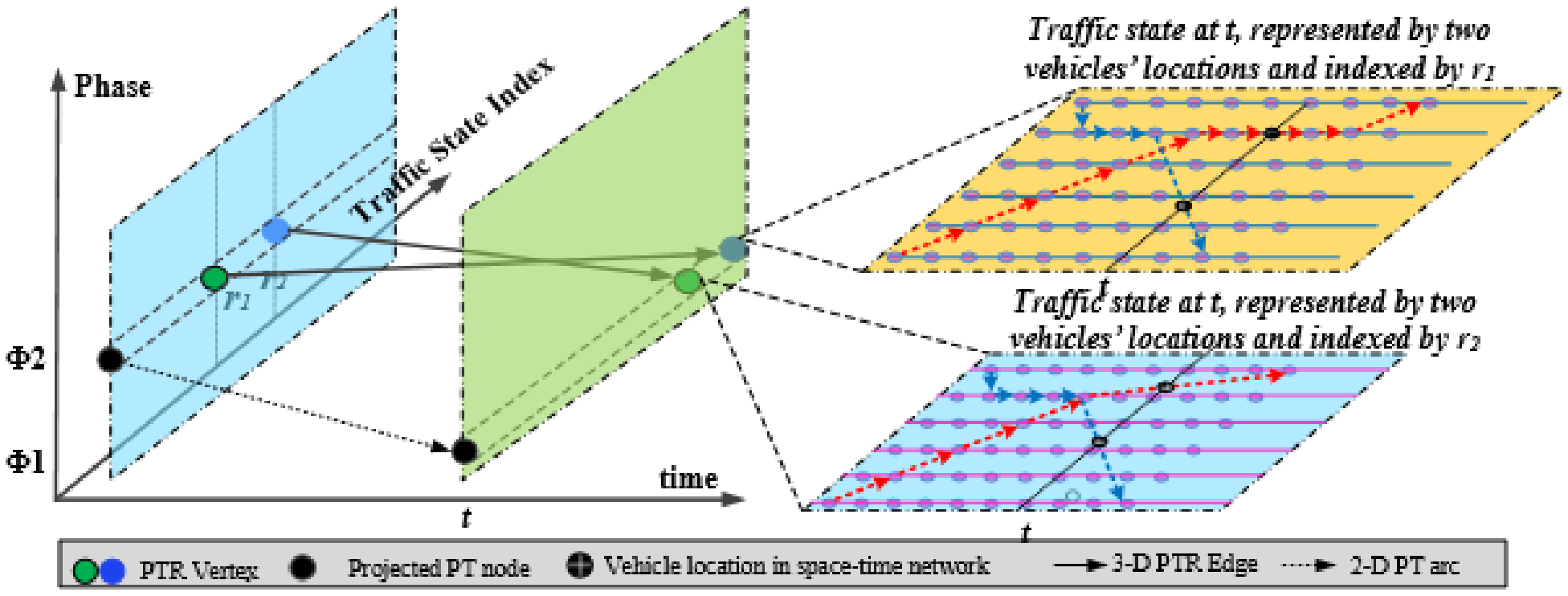

**Figure 4.** Illustration of the phase-time-traffic hypernetwork concept in which all vehicles, including CAVs and traditional vehicles, are released into the space-time plane and propagate. Interactions between vehicles and traffic signal operations can significantly affect vehicle trajectories [139].

Another group of AIM considers multiple intersections in a cooperative way [118][138][139]. In [138], a study is reported on how a single autonomous vehicle may cross multiple signalized intersections without stopping in a free flow mode; an optimal eco-driving algorithm is proposed to generate the acceleration and speed profile by considering multiple intersections jointly rather than dealing with them individually. The multi-intersection control is modeled using multiple agents across a network of interconnected intersections. From the multi-agent perspective, autonomous vehicles dynamically modify their scheduled paths based on different navigation policies and in response to minute-by-minute traffic conditions. Therefore, for a large road network, an instance of Braess' paradox may be experienced where opening additional travel options for the vehicles reduces the efficiency of all vehicles in the system. Such paradox is handled in [118]. Meanwhile, Li. et al. [139] have developed an intersection automation policy (IAP) for serving requests for green light made by both AV and human-driven vehicles. IAP exploits real-time tracking of vehicle location to predict arrival at intersections along its route where requests for green signals are anticipated. A schedule for green time is then devised based on the phase-time-traffic hypernetwork model, articulated in Figure 4, which represents heterogeneous traffic propagation under traffic signal operations. Thus, the signal time and vehicle movements are optimized for all vehicle types.

Like [139], other studies have considered intersection crossing by a mix of AVs and human-driven vehicles [140][141]. Assuming that the vehicle type can be determined, autonomous vehicles are safely directed through the intersection even if they arrive on a lane that is assigned a red signal. Scheduling platoon crossing an unregulated intersection is also one of the issues addressed in AIM. Generally, the problem is to schedule autonomous platoons through a *k*-way merge intersection; an intersection crossing involving two-way traffic has been shown as NP-complete [142]. A polynomial-time heuristic has been proposed for planning which platoons should wait so that others can go through in order to minimize the maximum delay for any vehicle. Recently, the scope of AIM work has been expanded to support the quality of the travel experience from the passenger perspective while still caring for trip safety and efficiency. Dai et al. [143] have proposed an intersection control algorithm that characterizes vehicular kinematic states and smoothness of the vehicle jitter, acceleration and expected velocity. The algorithm opts to alleviate the vehicle jitter by reducing sudden acceleration and deceleration and determine the right-of-way of vehicles by striking a balance between the traffic throughput and fairness among vehicles. As a result, the smoothness of vehicular movement has been enhanced and the travel time of individual vehicles is balanced by controlling the vehicle velocity.

### *3.2 Adaptive Traffic Light Control (ATLC)*

Adjusting the traffic signal timing is the most popular means for DTM [144]. Work that focuses on ATLC covers cases involving human-driven vehicles, autonomous vehicles, and a mix of both types. Existing schemes can be classified based on the scope and means for collecting real-time traffic data and the objectives of traffic control at an intersection.

<u>Data Collection:</u> Adapting the signal timing in real-time requires accurate assessment of the in-flow and out-flow at the intersection. To measure the arrival rate and queue length, some approaches have placed video cameras, piezo-electric sensors, and inductive loops at the intersection. These measurements are also used to anticipate changes in the traffic pattern and modify signal phases and timing accordingly. These DTM-based approaches have been shown to be quite effective in comparison to that of the pre-timed traffic light signals [145]. In addition, street-mounted sensor nodes also have been used to assess traffic to adjust green lights in order to improve the waiting time, number of stops, and vehicle density [146].

<u>Optimization Objectives:</u> The most common goal of ATLC is to maximize traffic throughput and minimize trip delay by adjusting signal phase and timing at an intersection. Some approaches additionally focus on reducing the total number of stops during the entire travel and thus ameliorating $CO_2$ emission. On the other hand, the focus of other work is on reducing the time and space complexity for solving the traffic signal control problem while improving the average waiting time [147]. ATLC has been further improved through anticipating traffic fluctuations at an intersection. Such anticipation is by correlating the real-time traffic data at an intersection $i_x$ with neighboring intersections. Such correlation is enabled by data sharing through various infrastructure-to- infrastructure (I2I) technologies. For example, in [145] a wireless sensor network deployed on the road network to assess upstream and downstream traffic density around $i_x$ to decrease the average waiting time of vehicles crossing an intersection $i_x$.

Using a broad range of real-time traffic data, a variety of methods to control traffic light signals have been proposed. For example, rule-based reinforcement learning ATLC is presented in [148], where the traffic lights of neighboring intersections coordinate locally; the work is extended by including an additional hierarchical observer/controller component at the regional level in order to better optimize the ATLC operation [149]. Moreover**,** multi-agent based algorithms have been applied to traffic light systems [150]-[156]; for instance multiple fuzzy logic controllers, interconnected using IEEE 802.15.4 technology are employed to dynamically order phases and calculate green time while factoring turns [150]. In addition, a distributed multi-agent system has been developed using sensors to monitor traffic volume variations. The system finds the shortest green period during a vehicle trip so that the experienced waiting time at intersections is minimized. Another group of multiple intersection ATLC algorithms exploit multi-agent reinforcement learning algorithms [153]-[157], where the reactions by local and nearby intersections are considered to adjust the traffic lights timing.

### *3.3 Dynamic Lane Grouping (DLG)*

Traffic light signal scheduling which generally opts to assign short green duration to less traffic demands; however, low in-flow still may occupy unnecessarily large number of lanes while vehicles pile up for making a turn. DLG algorithms opt to overcome such a limitation and increase the utilization of existing road resources by balancing between lane capacity supply and changes in turning demands. DLG algorithms have gained significant attention due to its adaptability to road capacity constraints. The main idea is to relieve traffic congestion and improve throughput and delay at intersections. The basic requirement for an effective DLG strategy is to estimate traffic volume for different movements, which cannot be provided by the inductive loop detection systems. As discussed in Section 2, traffic data can be collected and communicated various V2V, V2I, and I2I technologies such as road sensors, 802.11p, 802.16, i.e., WiMAX or cellular networks like LTE-V, or 3GPP Cellular-V2X(C-V2X). With the help of these advanced traffic monitoring technologies, one can estimate the vehicle count per turn at intersections.

Some studies have focused on the fundamentals of the DLG concept and formulated the problem mathematically. The popular objective is to maximize lane utilization at an isolated intersection under traffic demand variation. These studies define a maximum lane flow ratio as the assigned flow divided by the saturation rate; they strive to minimize changes in such saturation rate among different movements, which could lead to a significant performance degradation at intersections. The effectiveness of DLG has been demonstrated using numerical analysis compared to fixed lane grouping for varying number of lanes, saturated/unsaturated flow, and fixed/adaptive traffic signal timing. The performance is assessed in terms of the average delay at an intersection [97]-[104]. Some work has pursued DLG by combining two optimization problems, namely, signal timing control and dynamic road space allocation. In such a case the lane count and the possibility of having shared are considered as parameters in the traffic signal timing optimization. Figure 5 shows the layout of an example intersection for which the approach could be applied. The optimal lane group combinations and signal cycles are computed such that the average passing delay at the intersection is minimized [98]-[102].

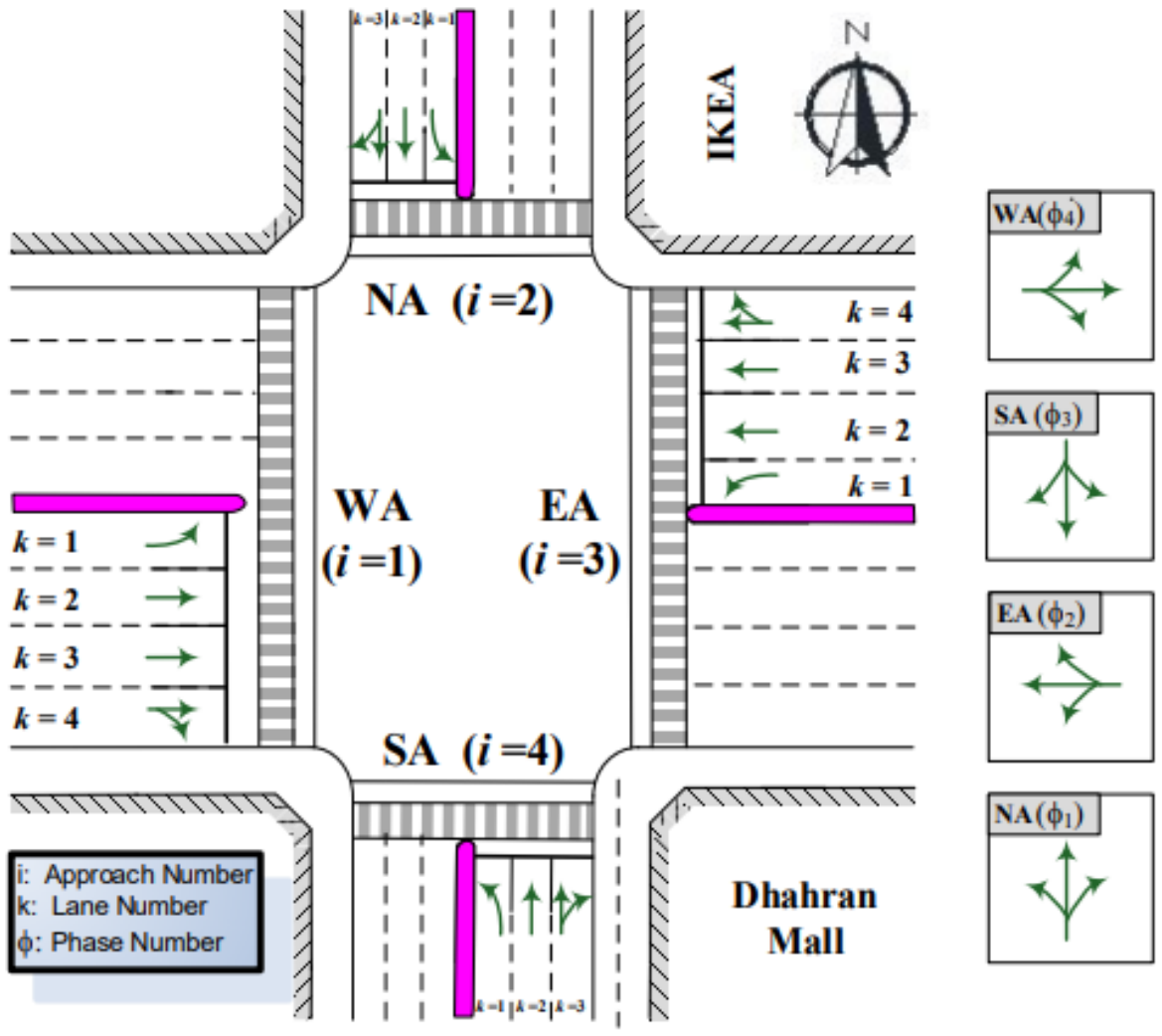

**Figure 5.** Example of intersection layout and turning movement assigned to each lane [98].

Evaluation based on case studies has been conducted by various approaches to validate the performance of DLG. Using microscopic traffic simulation, the benefits of a DLG strategy has been demonstrated and compared to the conventional fixed lane grouping in terms of mobility and sustainability [103]. The performance of DLG is assessed using the average vehicle delay, the number of stops during a trip, the average fuel consumption per vehicle, the average rate of pollutant emissions such as carbon monoxide (CO), hydrocarbons (HC), and oxides of nitrogen ($NO_x$) and $CO_2$ per vehicle. Finally, it has been shown that the impact of DLG grows as the traffic volume and the frequency of the turn pattern varies; a DLG strategy is effective in balancing lane flow ratios and reducing intersection crossing delay and consequently, energy and pollutants emission. Moreover, an automatic screening tool has been developed to identify the intersections for which DLG is advantageous [104]. Four assessment criteria are considered to evaluate traffic supply and demand, namely, (i) safe turning geometry which is a natural and logical prerequisite to qualify an intersection for DLG, (ii) volume change to measure traffic fluctuations between time periods, (iii) volume-to-lane (V/L) ratio and (iv) volume-to-capacity (V/C) ratio. The V/L and V/C capture the relationship between travel demand and capacity supply at an intersection. Through case studies, V/C has been shown to be the most effective criterion to identify a candidate intersection for DLG with a correct identification rate exceeding 90%; it has also been observed that DLG could reduce the overall intersection crossing delay by approximately 15% [104].

### *3.4 Dynamic Lane Reversal (DLR)*

Contraflow lane reversal is used in big cities during rush hours; such scheduling is clearly static and cannot cope with variation of the traffic intensity. Dynamic lane reversal (DLR) would be logistically complicated for traffic involving human-driven vehicles since the flow direction cannot be switched until the lane is empty; often the police has to be engaged to ensure that. With the emergence of CAVs, DLR is deemed to be a very viable option since autonomous vehicles can rapidly switch out of the designated lane for flow reversal due to the automatic motion control and the prompt V2I and V2V communication. A number of techniques have been proposed for DLR in CAV [118]-[121]; yet turns and collaborative intersection crossing are not addressed. In [118], DLR in collaboration with AIM has been proposed, where the total traffic volume on a road is monitored every two seconds; the road capacity is expanded by reversing the direction of a lane $r$ on the paired road $r_{dual}$ if the traffic demand on $r$ is 1.5 times larger than or equal to that of $r_{dual}$. Such work has demonstrated how CAV enables efficient utilization of road infrastructure. M. Duell et al. [119] have developed a DLR for increasing traffic flow on a congested downtown grid. Although their algorithm factors in dynamically changing traffic condition, it does not consider turns and their conflicts at an intersection. Levin et al. [120] also have focused on mitigating congestion of CAV traffic and formulated a DLR control problem for a single road segment as an integer program. A per-road agent is assumed for managing lanes and communicating with vehicles on the road. Meanwhile, Chu et al. [121] have considered mixed traffic scenario, yet only CAVs is assumed to travel on reversed lanes. The problem of optimizing schedules and routes on dynamically reversible lanes has been formulated as an integer linear program and evaluated using real-world transportation data.

Some DLR approaches have been specific to particular road layouts and cannot be generalized to other layouts [122]-[125]. The focus of Li et al. [122] is on a signalized intersection with six lanes and two additional reversible center lanes. Only four scenarios are considered for typical urban morning and evening peak-hours. In [123][124], a signalized diamond interchange is considered, where the proposed DLR approaches strive to handle the concern of space limitation for different turns in order to reduce oversaturation at the interchange. Krause et al. [123] opt to show the effectiveness of dynamic back-to-back reversible left-turn in collaboration with a traffic light signal (TLS) controller, while Zhao et al. [124] strive to maximize the reserved capacity of the internal lanes at the intersection considering a fixed set of TLS phases. On the other hand, the focus of [125] is on the applicability of DLR to exist lanes for dynamic left-turn traffic, as articulated in Figure 6. With the help of an additional traffic light (pre-signal) installed at the median opening, exit lanes for left-turn control problem was formulated as a mixed-integer nonlinear program, in which the geometric layout, main signal timing, and pre-signal timing were integrated and transformed into a series of mixed-integer linear programs. The results have shown significant growth in intersection capacity and reduction of traffic delay, especially under high left-turn demand.

Adjusting lane direction has been employed as an efficient way to overcome the logistical challenge in handling massive vehicular traffic during evacuation, where people are enabled to safely travel away from a hazardous site [106]-[114]. Some approaches like [106]-[108] use pre-known incoming traffic volume, and road capacities to find the optimal contraflow network configuration that minimizes the evacuation time. Then lane reversal is usually scheduled once and at the beginning of evacuation. Other studies have expanded the scope of the lane reversal optimization to consider additional factors such as evacuation priority for moving injured people to a hospital or transit priority for low-mobility people in large cities [109][110]. In addition, partial lane reversal is considered in [111]-[114] where only a subset of the lanes in certain directions are reversed to enable timely handling of evacuees who need urgent care, e.g., elders.

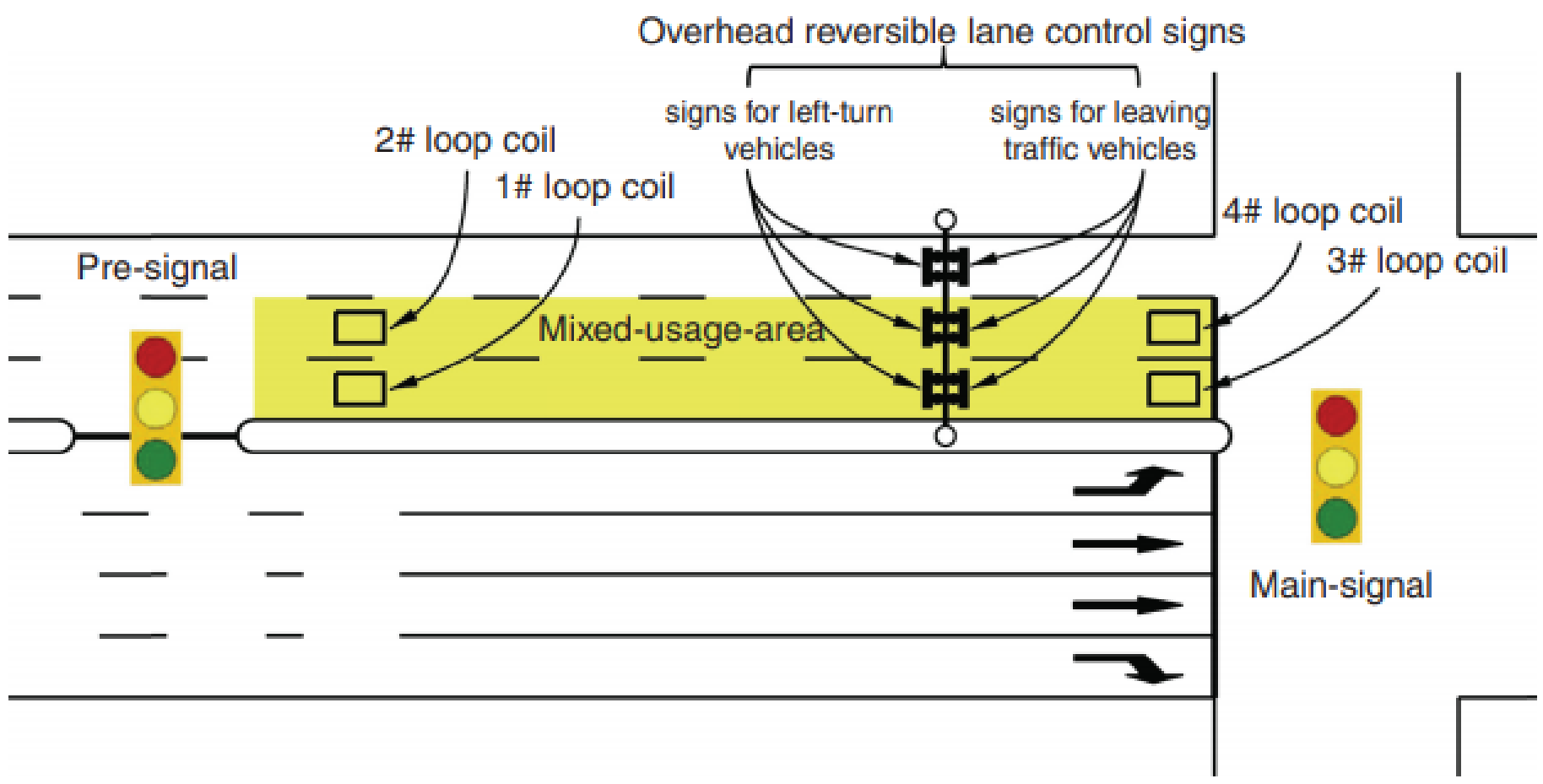

**Figure 6.** Exit lanes for left-turn traffic are controlled with a collaboration of geometric layout, main signal timing, and pre-signal timing. It shows location of reversible lane control signs and vehicle detectors. [125].

Another prominent application of lane reversal is to manage fluctuated volume of vehicular traffic in a particular direction depending on time or events in urban areas [115]-[117]. The objective is mainly to overcome the road capacity limitation in order to boost the vehicular throughput and reduce travel time. The focus of the work in this category is on determining the appropriate time to reverse a lane by monitoring traffic variations in both travel directions on a road segment. For instance, Zhou et al. [115] have developed a self-learning contraflow lane system for controlling tunnel traffic in order to estimate real-time traffic demand for passing the tunnel and decide when to use contraflow for preventing traffic jams. Meanwhile, M. Hausknechat et al. [116] have studied the impact of reversing a lane on an unsaturated road on vehicular throughput. They model the maximization of network traffic as a multi-commodity flow problem and propose a two-level formulation to calculate the optimal lane reversal configuration. Both approaches do not consider dynamically changing traffic demand and ignore the complication caused by turns at an intersection. On the other hand, T. Lu et al. [117] pursue a two-tier system for optimizing the reversible lane assignment while considering TLS settings and real-time traffic volume; the higher tier is for optimizing reversible lane assignment based on the total queue length at junctions while the lower tier is for traffic allocation at road segments.

### *3.5 Dynamic Trajectory Planning (DTP)*

The vehicular traffic assignment for the traditional human-driven vehicles is mainly derived by the stochastic nature of the problem that it is subject to uncertainties related to perception and reaction times of drivers and human based error. Such uncertainty can be mitigated in CAV. Nonetheless, CAV raises new issues given the fine-grained controllability of autonomous vehicles. Basically, the spacing between autonomous vehicles can be significantly reduced enabling a set of vehicles to travel as a platoon. Thus, forming a platoon, joining and departing of an existing platoon, and setting the appropriate vehicle configuration are unique challenges in the case of CAV. In essence, the vehicles have to collectively determine speed, acceleration, and optimal spacing between them, subject to safety and road condition constraints. Therefore, DTM will not only have to optimally assign traffic but also have to find the optimal vehicle trajectories. Indeed, the path selected or assigned by routing models in DTM will be subject to the multi-CAV motion planning for autonomous or mixed traffic scenarios.

CAV motion planning is categorized in [158][159] into four hierarchical classes: (1) route planning that aims to find the best global route for given O-D pairs and corresponds exactly to the traditional traffic assignment. The individual vehicular route is derived based on traffic statistics (current and anticipated) and does not consider obstacles, road geometry, etc.; (2) path planning which is a bit more fine-grained and considers the constraints of individual road segments that connect the origin and the destination and opts to cope with obstacles and specific flow constraints, e.g., due to construction; (3) maneuver planning which determines the appropriate decision in each step including for example ‘going straight’, ‘going left’, etc. It considers further the position and speed of the CAV while taking into account the path that is specified from path planning; (4) trajectory planning which governs the motion of the vehicle and determines the vehicle’s transition from one feasible state to another while considering road obstacles and the vehicle’s kinematic. In essence, trajectory planning is concerned with the vehicle control.

In the realm of CAV, dynamic traffic assignment constitutes autonomous motion planning to find a feasible route over collision-free path towards destination while taking into account the vehicle dynamics and maneuver capabilities, and respecting the traffic rules and road boundaries [159][160]. The motion planning algorithm would utilize sensor readings and supplement them with data from digital road maps in order to provide a sequence of state transitions for the vehicle controller. In other words, the CAV will iterate over set of states and actions to implement the plan defined by DTM. The planning problem is quite complicated given the search space and the number of optimization parameters. Many map tessellation solutions have been proposed in the literature in order to tackle such complexity [159]. Search space reduction techniques can be categorized as local or incremental search, depending on the existence of possible prior states. Once the feasible positions are determined, a continuous interaction between path planning and maneuver selection is used in order to ensure respecting road rules and avoiding obstacles. Such interaction is detailed in [159]. It should be noted that such high-granularity for motion planning is not warranted for human-driven vehicles due to the presence of a driver. Given the scope of the chapter, we will focus mainly on the individual trajectory optimization, collaborative trajectory optimization and stream-based optimization, and reconfiguration in homogeneous and heterogeneous setup.

Individual trajectory optimization: The objective in this category is to minimize the length of the trajectory, its smoothness and the offset from the central line of the lane, which is used as a reference path, under constraints imposed by the routes [161]. Other metrics include safety [162][163], intersection crossing efficiency [164], and fuel consumption [165]. Multi-objective trajectory optimization has also been considered; Ma et al. [166] promote trajectories that simultaneously optimize travel time and fuel consumption for all vehicles. The optimization constraints usually include the road boundaries, lane restriction, trajectories curvature, speed limit, acceleration rate, and obstacles. The obstacles are generally represented as circles to avoid collision using colliding trajectories detections. Some work also considers maneuvering and traffic rules [167] and generates trajectories that respect the checkpoints determined by path planning, stop signs, traffic lights, turns, lane changes, intersection crossings, turns, and dead-ends. The trajectory efficiency is measured using the distance and time until reaching the next checkpoint and the number of possible collisions with obstacles. The complexity of determining optimal trajectory is very high; some approaches apply machine learning to infer trajectory patterns from human driving [168] and then exploit them for online trajectory generation.

Coordinated stream of vehicles: Coordination among a stream of vehicles has been studied by optimizing the trajectory of multiple CAV. Such a problem is very challenging as the vehicles are constrained by car-following models and each platoon have some characteristics like acceleration, speed, and safety distance. Two types of optimization have been considered: (1) how to optimize the traffic assignment to avoid frequent switching of traffic lights and the frequency of vehicle stop/start, which will negatively affect the travel time. This category is covered in Section 3.1 of this chapter; (2) how to minimize the traffic flow fluctuation by choosing the appropriate configuration for each vehicle in a platoon or the speed of the leading vehicle in order to improve throughput and other performance metrics. In [169], the problem is formulated as a mixed integer program and used dynamic programming to solve it. Others, e.g., [166][170], proposed a shooting heuristic that can effectively smooth the trajectory of a stream of vehicles approaching a signalized intersection by detailed control of their acceleration profiles. The shooting heuristic reduces the complexity by representing the trajectory search space as a few segments of analytical quadratic curves. The trajectories are constrained by the vehicle physical capabilities, safe inter-vehicle spacing, and traffic signal timing. However, only fixed signal timing and phasing are considered to control vehicle trajectories. On the other hand, some approaches reserves a certain lane for intersection crossing without considering any explicit traffic light [171][172].

Optimization in heterogeneous setups: The coexistence of human-driven and autonomous vehicles limits the ability of CAV to improve traffic performance using ramp metering, variable speed limits, signal control, etc. This is due to the traffic disturbance caused by drivers which could force cooperative lane changes, or unwarranted variability in inter-vehicle spacing, stop-and-go triggered by rubbernecking. To mitigate such disturbance, CAV longitude control has been commonly used, specifically, by maintaining a constant spacing or headway (or time) between successive vehicles. To avoid collisions within a platoon, the CAV controllers have to be designed to ensure string stability. There exist three major approaches for CAV longitudinal control [173]: linear, optimal such as model predictive control, and artificial intelligence (AI) based. Linear controllers focus mainly on string stability by determining the appropriate feedback and feedforward gains to adjust the acceleration. The optimal control approach uses multi-objective formulation that ensures control efficiency (e.g., relative speed to a platoon leader) and comfort (e.g., acceleration) [174]. Motivated by the fact that self-driving can be seen as data driven learning, AI-based controllers exploit machine learning techniques instead of parametric rule-based models [175]. Finally, the adoption of electric vehicles is another type heterogeneity given that they have greater range of acceleration/deceleration [173][176]. Another source of heterogeneity is user-customized CAVs, where the CAV behavior is influenced by human preference in terms of desired speed and/or acceleration rates, etc. [173][177].

**Table 2:** Summary of traffic management algorithms;
AV: autonomous vehicle, HV: human-driven vehicles, SLR: static lane reversal

| **Reference** | **Category** | **Sub-category** | **objectives** | **Intersection** | **Junctions** | **Vehicle Type** |
|---|---|---|---|---|---|---|
| [97] | DLG | Optimize space only | Minimize fluctuation of a maximum lane flow ratio, and reduce average delay | Signalized | Single or multiple | HV |
| [98] | | Optimize space and time allocations | Minimize average intersection crossing delay | | Single | |
| [99] | | | Minimize delay | | | |
| [100] | | | | | | |
| [101] | | | | | | |
| [102] | | | | | | |
| [103] | | | | | | |
| [104] | | | Reduce delay; increase throughput | | Multiple | |
| [105] | | | Increase capacity | | Single | AV |
| [106] | DLR | Optimization of lane-based evacuation route | Minimize evacuation time | No signal | Multiple | HV |
| [107] | | | | | | |
| [108] | | | | | | |
| [109] | | | | | | |
| [110] | | | | | | |
| [111] | | Optimize partial lane reversal | Reduce evacuation/clear time | | | |
| [112] | | | Prioritize evacuees with urgent care, e.g., elders | | | |
| [113] | | | Reduce evacuation time | | | |
| [114] | | | | | | |
| [115] | | Optimize tunnel lane reversal | Increase accuracy of traffic demand prediction | | Tunnel | |
| [116] | | Optimize traffic flow | Increase road network efficiency | | Single | |
| [117] | | Optimize traffic assignment | Reduce travel time | Signalized | | |
| [119] | | Optimize route selection | Minimize total travel time | No signal | | AV |
| [120] | | Optimize traffic assignment | Maximize vehicular flow | | Single | |
| [121] | | | Optimize travel schedule | | Single | |
| [122] | | | Optimize layout | signalized | Arterial roadways | HV |
| [123] | | Focus on left-turn | Reduce delay; increase throughput | | Diamond inter-changes | |
| [124] | | | Expand capacity; reduce congestion | | | |
| [125] | | Focus on left-turn | Increase capacity; reduce waiting delay | | single | |
| [126] | AIM | Centralized | Minimize total travel time | No signal | Multiple | AV |
| [127] | | | Minimize a sum of vehicle exit time | | Single | |
| [128] | | Decentralized | Minimize travel time with low budgets | Signalized | | |
| [131] | | Decentralized | Increase safety under the high rate of accidents | No signal | | Semi-AV |
| [132] | | Trajectory optimization | Decrease traffic delay | | | Semi-AV |
| [133] | | | improve safety; minimize stops and travel time | | | AV |
| [134] | | | Improve ratio of average trip time to throughput | | | |
| [135] | | | Maximize road capacity | | | |
| [136] | | | Optimize AV trajectories and signal control | Signalized | | Mixed |
| [137] | | | Improve the computational efficiency | | | AV |
| [129] | | Decentralized | Improve the computational efficiency | No signal | | |
| [130] | | | Minimize energy; maximize throughput | | | |
| [138] | | multi-intersection optimization | Minimize energy consumption and travel time | Signalized | Multiple | |
| [118] | AIM & DLR | | Minimize travel time | | | |
| [139] | AIM | Optimize traffic signal timing and vehicle movements | Minimize total delays | | | Mixed |
| [140] | AIM & DLG | Hybrid of trajectory and TLS | Minimize delay and maximize throughput | | Single | Mixed |
| [141] | AIM | Trajectory optimization | Balance and maximize traffic flow rate | No signal | | Mixed |
| [142] | | | Minimize the maximum delay for any vehicle | | | AV platoon |
| [143] | | | Improve fairness, safety and throughput | | Single | AV |
| [145] | ATLC | Optimize green time duration | Minimize average vehicle waiting time | Signalized | Multiple | HV |
| [146] | | Optimize sequence and length of traffic lights | Improve delay and throughput | | Single | |
| [147] | | Optimize signal phase and time | Reduce time and space complexity | | Single | |
| [148] | | Simulation based | Improve delay and throughput | | Multiple | |
| [149] | | Centralized/decentralized | Improve delay reducing fuel consumption | | | |
| [150] | | Optimize signal phase and green time duration | Improve delay and throughput | | | |
| [151] | | Optimize signal phase and green time duration | Minimize average vehicle waiting time | | Single | |
| [152] | | adjust green time duration | Reduce average vehicle waiting time | | | |
| [153] | | Optimize signal phase and green time duration | Improve average travel time | | Single or multiple | |
| [154] | | Optimize signal phase and green time duration | Improve delay and throughput | | Multiple | |
| [155] | | Coordinated traffic signal control | Reduce total number of stopped vehicles | | Single | |
| [156] | | Reduce computational complexity | Improve delay and throughput | | Multiple | |
| [157] | | Optimize signal phase and green time duration | Improve delay and throughput | | | |
| [161] | DTP | Optimize individual trajectory | Improves safety and vehicular throughput | No signal | N/A | AV |
| [162] | | Optimize individual trajectory | Increase Safety | No signal | N/A | AV |
| [164] | | Multi-trajectory optimization | Minimize travel time (highway lanes) | No signal | N/A | AV platoon |
| [165] | | Optimize individual trajectory | Improve fuel efficiency | Signalized | N/A | AV |
| [166] | | Coordinated stream of vehicles | Optimize delay, energy and safety. | Signalized | N/A | AV |
| [167] | | Coordinated stream of vehicles | Increase Safety | Signalized | N/A | AV |
| [168] | | Optimize individual trajectory | Reduce computation complexity | No signal | N/A | AV |
| [169] | | Coordinated stream of vehicles | Improve safety and throughput | No signal | N/A | AV |
| [170] | | Coordinated stream of vehicles | Improve safety | Signalized | N/A | AV |
| [171] | | Coordinated stream of vehicles | Improves safety and vehicular throughput | N/A | N/A | AV |
| [172] | | Coordinated motion of vehicles | Reduce travel time; increase safety | No signal | N/A | AV |
| [173] | | Handle heterogeneous setups | Analyze safety | No signal | N/A | AV and HV |
| [174] | | Handle heterogeneous setups | Improve safety | No signal | N/A | AV |
| [175] | | Handle heterogeneous setups | Improve safety and fuel consumption | No signal | N/A | AV |
| [176] | | Handle heterogeneous setups | Improve safety | No signal | N/A | AV |
| [177] | | Handle heterogeneous setups | Risk analysis | No signal | N/A | AV |

# 4 Smart Road Vision and Practical Issues

As pointed out in the previous section, CAV will revolutionize the transportation industry and will enable effective management of road traffic to achieve optimal performance. Yet, the reality is that human-driven vehicles will not disappear anytime soon and numerous practical issues that ought to be considered. In this section, we highlight these issues and present our iRoad vision. The iRoad project, which stands for Internet of Radio-equipped On-road and Vehicles-carried Agile Devices, opts to tackle the challenges in realizing the DTM methodology when a mix of autonomous and human-driven vehicles share the road.

## *4.1 Support of Human-driven Vehicles*

CAV provides three key features that facilitate the implementation of DTM. First, the behavior of an autonomous vehicle is predictable; meaning that if instructed to follow a certain route, or even, trajectory, it indeed does so. Thus, one can estimate the vehicle density on the various road segments with high fidelity. In other words, the data collection is quite easy and accurate when only autonomous vehicles are on the road. Second, each autonomous vehicle is capable of wireless communication. Such capability will enable instructing a vehicle about route change, sending alerts, and receiving road and vehicle status updates. The third feature is the ability of an autonomous vehicle to precisely and safely maneuver and follow a prescribed trajectory at a fine-grained level. Obviously these features are not available for human-driven vehicles. Thus, when both autonomous and human-driven vehicles exist on the road, the realization of DTM will be quite challenging and the idealistic view about vehicle compliance would be unrealistic. In other words, human behavior and uncertainty about vehicle navigation and status complicate DTM immensely.

VANETs have been explored as a means to enable DTM for traffic involving contemporary human-driven vehicles [178]. In a VANET, cars are nodes that collaborate in assessing road conditions and sharing safety information [179]. Every participating vehicle also acts as a wireless router to allow nearby cars to connect with each other, creating network topologies that span vast areas [180]. Attempts to deal with congestion using VANETs have been made, mainly by using VANETs to discover and disseminate congestion information [181], providing routing recommendations [182][183], and setting traffic signal timing [25][63]-[65]. We argue that a VANET that alerts drivers about accidents or traffic bottlenecks would be ineffective in resolving congestion. In addition, routing around bottlenecks and scheduling a traffic light at an intersection ease congestion rather than reduce the probability that it indeed occurs. Moreover, existing VANET-based techniques assume appropriately equipped vehicles, which hinders their applicability since most vehicles on the road today do not have onboard wireless transceivers.

Our iRoad project opts to overcome these limitations and promotes a novel and cost-effective system that can dynamically adapt to traffic conditions and operate autonomously without the need for costly on-road sensors. The iRoad system employs driver-carried smartphones and tablets, and leverages the popularity of peer-to-peer (P2P) networking to optimize both road and driver centric metrics. The overall methodology can be viewed as managing vehicular traffic using participatory-sensing by the drivers and their vehicles. Unlike existing approaches, e.g., Google maps, Waze [27], Inrix [28], Cellint [29], etc., the goal is to reduce congestion by balancing the load on local roads through, (1) predicting increased traffic, (2) accounting for the impact of driver's route choices, and (3) providing route recommendations while factoring in potential changes in the capacity of certain road segments. Every vehicle chooses a route to its destination and generates a set of optional routes. We have developed a P2P system to be used on each vehicle to query other vehicles in its neighborhood on whether they are travelling on any of the edges of its main and optional routes. Based on data collected from traffic reports, alerts, and vehicles, our system would make a decision for the vehicle to continue on the same route or suggest an alternate route. The combined effect of all vehicles would impact the traffic pattern and allow smooth and faster travel for everyone. In other words, our system opts to achieve user equilibrium. Our preliminary results showed that 20% reduction in travel time could be achieved by such an approach even if only 40% of the drivers follow the route recommendations [32].

Smart road configuration capabilities, e.g., making a lane HOV or switching its direction, are also being exploited as a means for influencing vehicular flow on a particular road. Thus, our iRoad system achieves the DTM goals by providing real-time traffic management and being proactive in preventing traffic congestion [5]. Figure 2 shows a functional diagram. Our system has three major players: the vehicles, the VANET, and the smart road configuration controller. The vehicle part constitutes the user, i.e., driver interface in case of human-driven vehicles, and the route selection and action planning algorithm. The realization of the vehicle part is as either an app for a smart device or software module that is integrated in the vehicle's dashboard panel. Every vehicle $V_i$ has a unique identifier, e.g. VIN number, and an onboard GPS receiver (smart phone). Neighbors of a node (vehicle) are those it can directly reach through P2P links.

Our system can be implemented using a centralized traffic management controller, where a local server aggregates the data, configures the roads, and provides route recommendations; Figure 3 shows the system architecture. Moreover, the system can also be realized in a fully distributed manner where a vehicle (user's smartphone) factors in the data collected in its vicinity in route selection and local road configuration controllers, e.g., traffic light controllers, adjust their operation to improve traffic flow and the utilization of the road capacity. All computations for route selection are to be performed on the vehicle itself, e.g., using a smart portable device in case of human-driven vehicles. In such as a case, the driver is to choose a destination and travel route. For autonomous vehicles, part of this functionality such as the route selection, could be performed by a road-side unit or a remote traffic management system to optimize some global metrics. While some vehicles may not join the network due to lack of equipment or due to the driver's preference, with the right incentive the bulk will join a VANET on the road. Incentives could simply better user experience or reduced toll or vehicle registration fees. Although the system is completely autonomous, we expect drivers to decide on staying on the current route or re-routing to suggested paths whenever applicable.

We envision our system as laying the foundation for handling mixed traffic of autonomous and human-driven vehicles where coordination among these diverse sets of vehicles and between them and smart road controllers will be expected and/or mandated. The iRoad system can also facilitate the handling of major evacuation scenarios where traffic can be chaotic. In cases of predicted disasters, such as hurricanes, massive evacuations are often necessary [188]. Our system will enable large scale sharing of critical information without reliance on the communication infrastructure, which could have been degraded by the disaster, and also enable configuring roads autonomously as needed in order to support organized evacuation. To realize such a system, our research group is tackling the following key tasks. Later in the section we summarize some of the results; more can be found in [32][189]-[194].

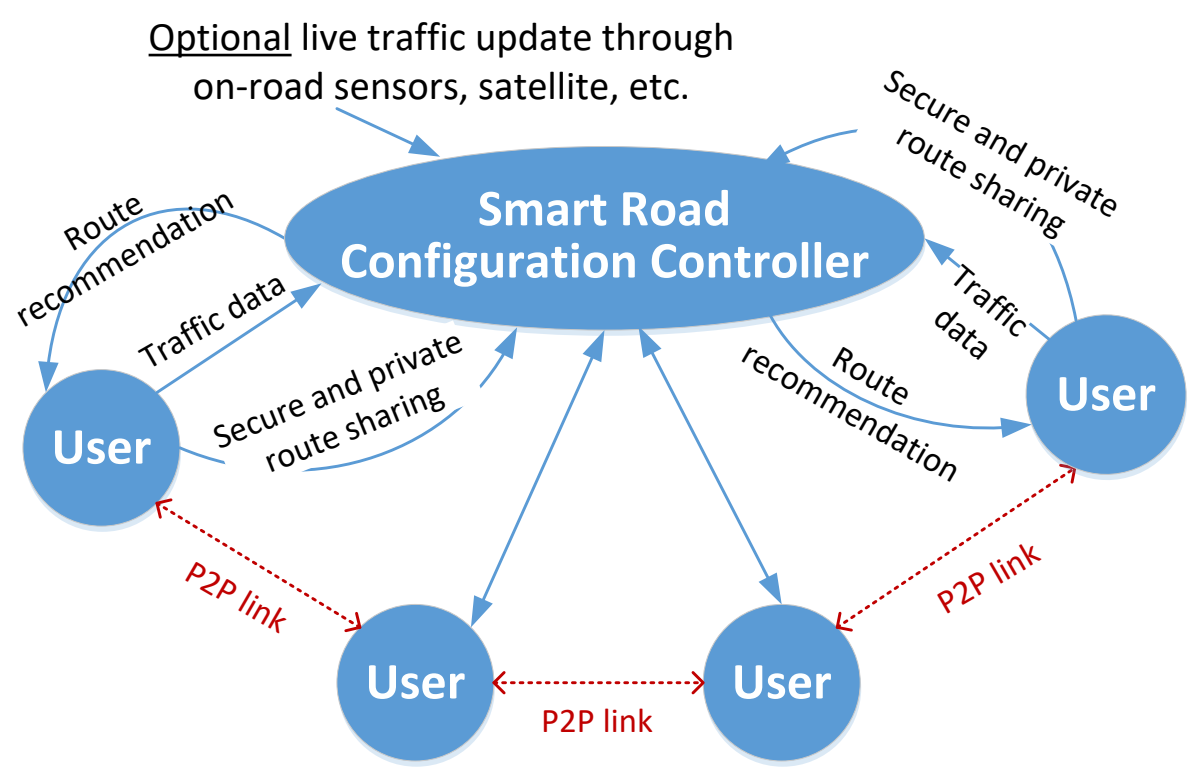

**Figure 2**: A functional diagram of the proposed system. Users may be drivers of contemporary vehicles or autonomous vehicles. Users securely share their route plans, while factoring in traffic data, collected by participatory sensing and optionally, using on-road sensors and live traffic monitoring facilities when available. The smart road configuration controller (SRCC) may be a regional server or a road-side unit. Route recommendations can be provided by the SRCC, or be determined in a distributed manner, i.e., vehicles collaboratively manage their travel paths and implicitly influence road configuration.

1) *Developing novel routing schemes for preventing congestion and slow traffic* – The key objective is to balance the load on the roads and prevent congestion through inter-vehicle data sharing while factoring in traffic data that is made available by authorities. Both human-driven and autonomous vehicles are assumed. The routes planned by other vehicles influence a vehicle's own decision, implying that the system takes advantage of the highly dynamic VANET to make a fine-grained analysis and yield a more optimized travel route. Issues related to uncertainty about the collected data from other vehicles are also being addressed.

2) *Developing a P2P system for efficient internetworking of collocated devices* – Despite the growing push for VANET technologies, practically only a small fraction of the vehicles on the road have modems and autonomous vehicles are expected to stay a minority in the near future. Therefore, alternative means for inter-vehicle data sharing is needed. We have developed a P2P system that operates on smart cell phones and tablets to collect and share sensor data among vehicles, and to communicate with road configuration controllers. Specifically, we use Wi-Fi Direct to enable internetworking of these devices and extend the protocol stack, and build software libraries for the realization of our system. Our approach not only enables real-time data sharing but also expedites the integration of other prominent technologies like DSRC by providing means for old vehicles to participate.

3) *Developing adaptive strategies for road configuration to effectively react to traffic problems* – We exploit existing facilities for changing lane specification (e.g., switching traffic direction, designating a lane as HOV, adapting toll assessments, etc.) to enable major traffic flow optimization. We opt to devise optimization models for flow maximization while factoring in possible impacts, e.g., lost toll revenue.

4) *Developing a protocol to enable sharing of route information while protecting drivers' location privacy* – One of the obvious concerns in sharing routes and other observed/sensed road conditions is how to sustain the privacy of drivers. In addition, knowing the future location of vehicles can be misused by attackers to conduct physical attacks and robbery.

5) *Developing schemes to guard against contemporary security attacks* – The autonomous nature of the system and the participatory nature of the collected data raise the concern about various attacks launched to disrupt its operation and/or to unjustly gain privileges on the road. An attacker could pretend to be multiple simultaneous vehicles at different locations [195] in order to influence other vehicles routing decisions. The attacked can also inject false routes for many non-existing vehicles to create an illusion of traffic congestion in order to trigger a favorable road reconfiguration [196].

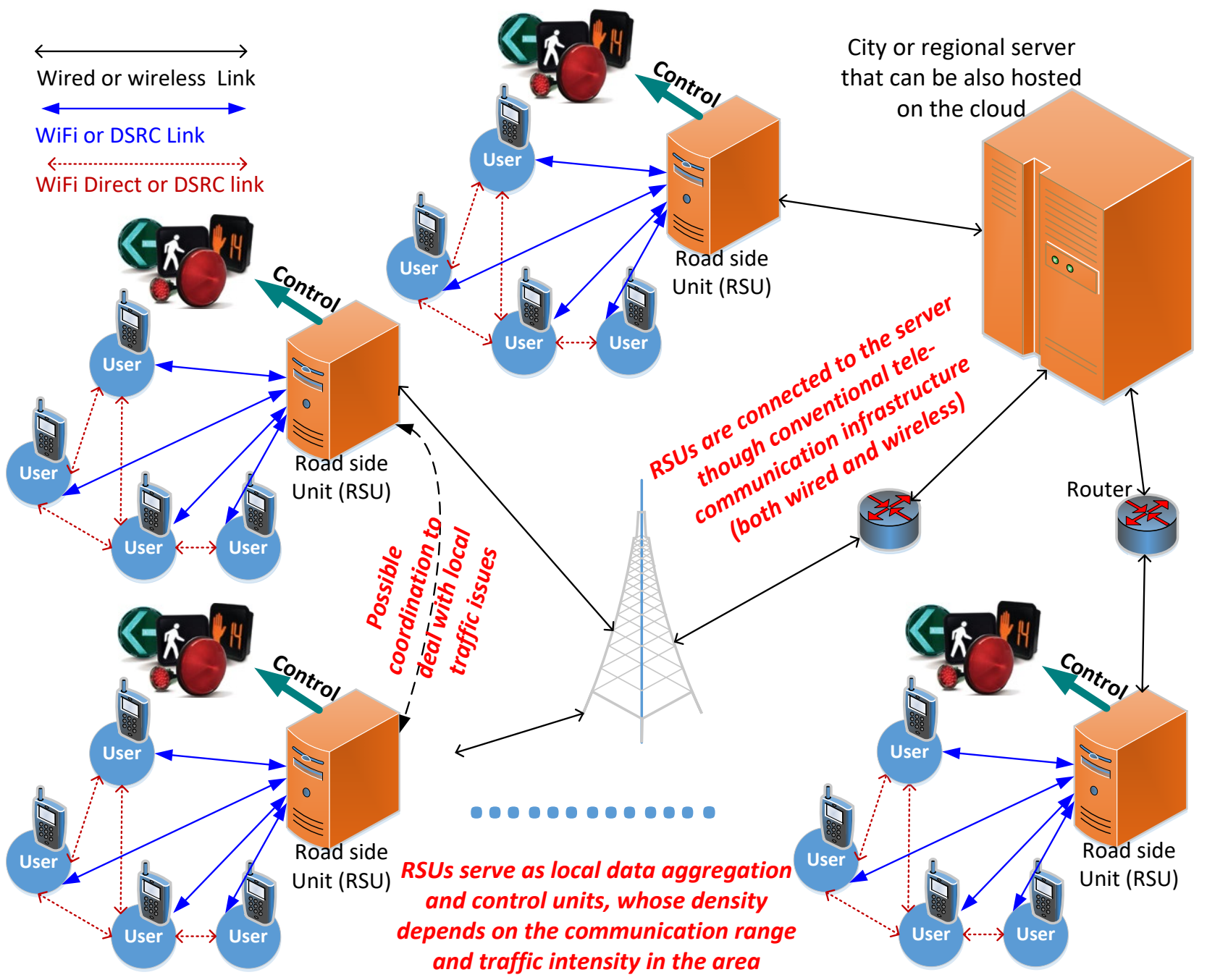

**Figure 3**: Realization of the proposed system, where traffic is autonomously managed locally, e.g., at the level of city downtown area, by road side units, and regionally, e.g., for a large metropolitan or county, by a server possibly while involving authorities for major road configuration decisions. Participatory sensing enables an economic tracking of road conditions and makes drivers part of the traffic management optimizations.

### *4.2* *Optimized Route Selection in Mixed Traffic*

Due to the unpredictably varying nature of the vehicular traffic, it is hard to accurately foresee the number of vehicles travelling on the roads in order to avoid congestion. As pointed out above, existing systems opt to mitigate the traffic uncertainty by providing the current status, alerting drivers and providing route recommendations. However, that would not suffice since the response of the individual drivers varies, and the rerouting decision made by them may cause congestion somewhere else. Our iRoad system not only provides for an increase in situational awareness but also enables exchanging routing plans among the drivers, and with SRCCs to improve the vehicle throughput and travel time. Thus, the routes that others take would influence a vehicle's own decision and potentially affect the traffic pattern on certain roads. Such a decision-making process enables even distribution of vehicular traffic since one can predict the condition of alternative routes before changing the travel path. However, these features come at the expense of increased communication and computation overhead. In addition to the inter-vehicle messaging for data sharing, the routing decision under our model will be more complex than typical. An accurate forecast of traffic intensity is necessary to assure drivers that it would be advantageous to use different routes. Moreover, the effect of potential road reconfiguration has to be considered. While changing the road configuration is an effective means for congestion mitigation, it complicates the routing problem. Basically, the time taken to travel a particular segment will vary due not only to traffic volume but also to changes in road capacities. Such assurance and positive experiences would increase vehicle participation.

To assess the potential impact of sharing travel routes among vehicles and dynamic road configuration, specifically adjusting traffic light timing on performance, we have conducted a preliminary study [32]. The study is based on the following operation model. All vehicles have the same communication range $D_R$, which is to be carefully set to reduce interference, especially in congested areas. When a vehicle has to retrieve information from others in the vicinity, it broadcasts a message with its location, routes and speed; all vehicles receiving this message are expected to reciprocate by responding with their information. In addition to the route $R_C$ that $V_i$ currently follows, every vehicle has a predefined set of *k* optional routes $R_O[i]$, *for* $i=\{1, \ldots k\}$ [182]. Figure 4 shows an articulation of a possible configuration. Every interval of time ($I_T$), a vehicle broadcasts a message that contains its ID and the next "*N*" edges on route $R_C$, as well as all edges on optional routes. The value of *N* is pre-determined and is the same for all vehicles. Based on the received updates from neighbors, a node estimates the number of vehicles on the edges of $R_C$ and all $R_O[i]$ and the expected delay based on the vehicle count per unit distance for these edges. Using the delay estimate, the node assesses the travel time by factoring in the waiting time at the traffic lights that will be encountered and decides whether the current route is still the best choice or one of the optional routes would be better. The driver may follow such a recommendation with a probability $P_R$. On the other hand, an ALTC will use the data to adjust green time. The road with more vehicles would get more green time and vice-versa. No inter-ATLC coordination is considered.

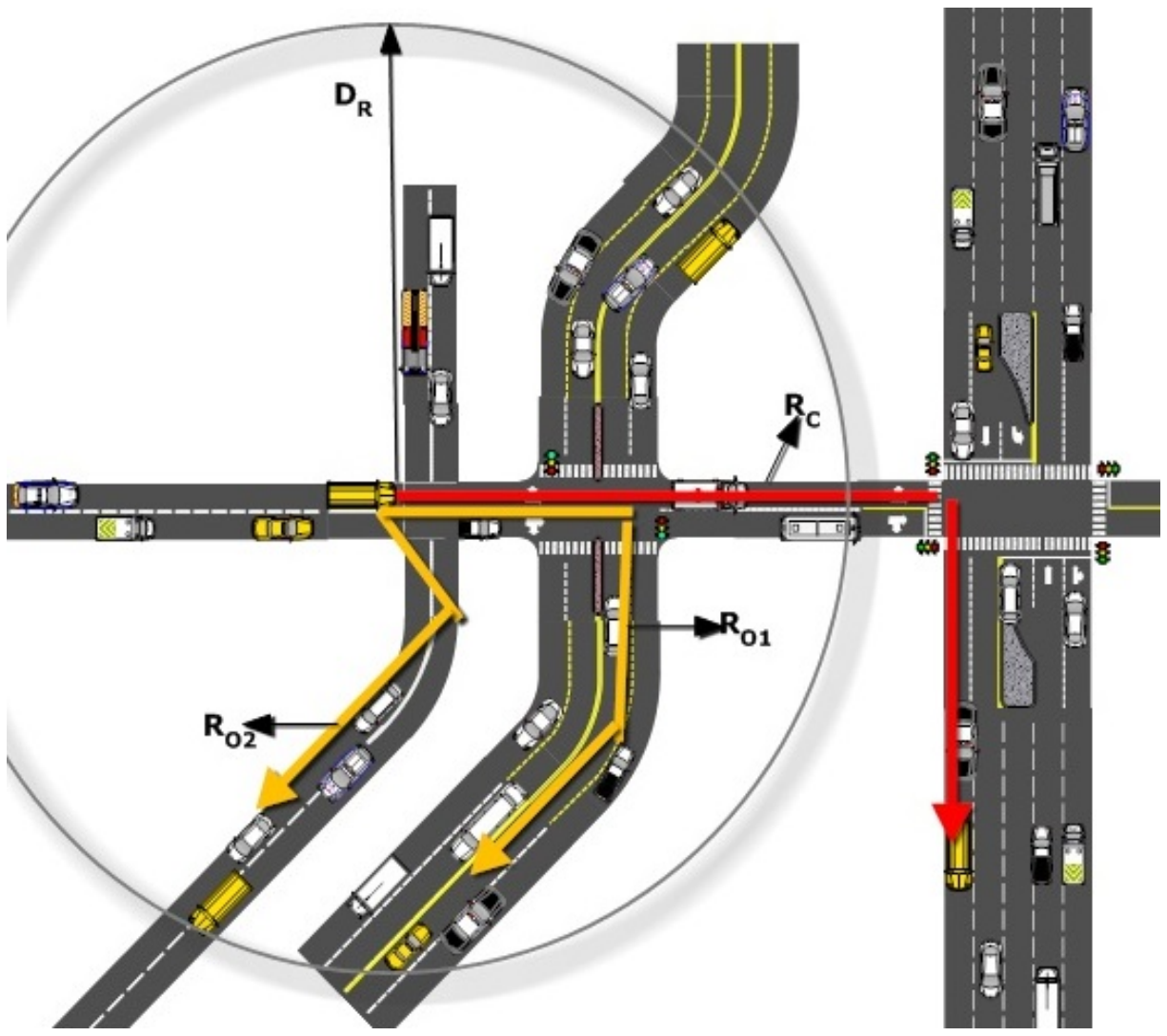

**Figure 4**: Illustration of the assumed system model showing the relevant vehicle parameters.

We have simulated such an operation model for a 15 × 15 km$^2$ area in downtown Baltimore. A total of 4,000 vehicles are allowed to travel in that area. For each vehicle, the departure time is randomly selected between 0 and 1,000, and the origin-destination pair is also chosen randomly within the area. For fair comparison, we use the same set of vehicles with the same departure times and origin-destination pairs as input with and without VANET-enabled Autonomous traffic Management (VAM). We performed 30 simulation runs, implying 30 different origin-destination pairs, and averaged the results. In the simulation, $D_R$, $I_T$, *N*, and $P_R$, were set to 5 km, 5 units, 5 edges, and 0.85, respectively. We introduced congestion by forcefully stopping two specific vehicles, for 30 simulation time units at predefined times when they enter specific edges. Given the goal of the study, we ignored delays due to message queuing and retransmission. We compare the performance of VAM to two other traffic management mechanisms. The first is a centralized routing scheme, similar to that described in [44], which bases the routing decision on the state of the entire set of roads. Obviously, in practice, this approach entails massive state updates. Nonetheless, it would serve as an upper bound for how well the vehicular traffic can be managed. The second baseline approach is called *Autonomos* [197], which alerts vehicles entering a congestion zone to avoid vehicle pile-up. We also compare the performance to the case when no action is taken to mitigate traffic congestion.

Figure 5 shows the time until all the simulated vehicles reach their destination. The graph shows that VAM, which assigns routes in real time and relies only on the local state, almost matches the performance of a centralized approach. *Autonomos* shortens the trip time since vehicles receive an alert message when they come close to a congested area, which could be used as a signal to re-route. Yet, *Autonomos* cannot prevent the occurrence of congestion and yields significantly worse results than VAM. The gap between VAM and *Autonomos* grows with the increase in the vehicle population. The plot for the conventional routing shows non-uniformity as compared to the other approaches, since it does not deal with congestion. It has a high number of vehicles on the roads in the middle of the simulation and thus the delay grows. At the beginning and towards the end of the simulation, the number of vehicles that is still on the roads is fewer and the delay lessens. All the other approaches have provisions for dealing with congestion and hence have a uniformly rising curve. The results have demonstrated a major improvement even with such a scaled down and primitive version of our methodology.

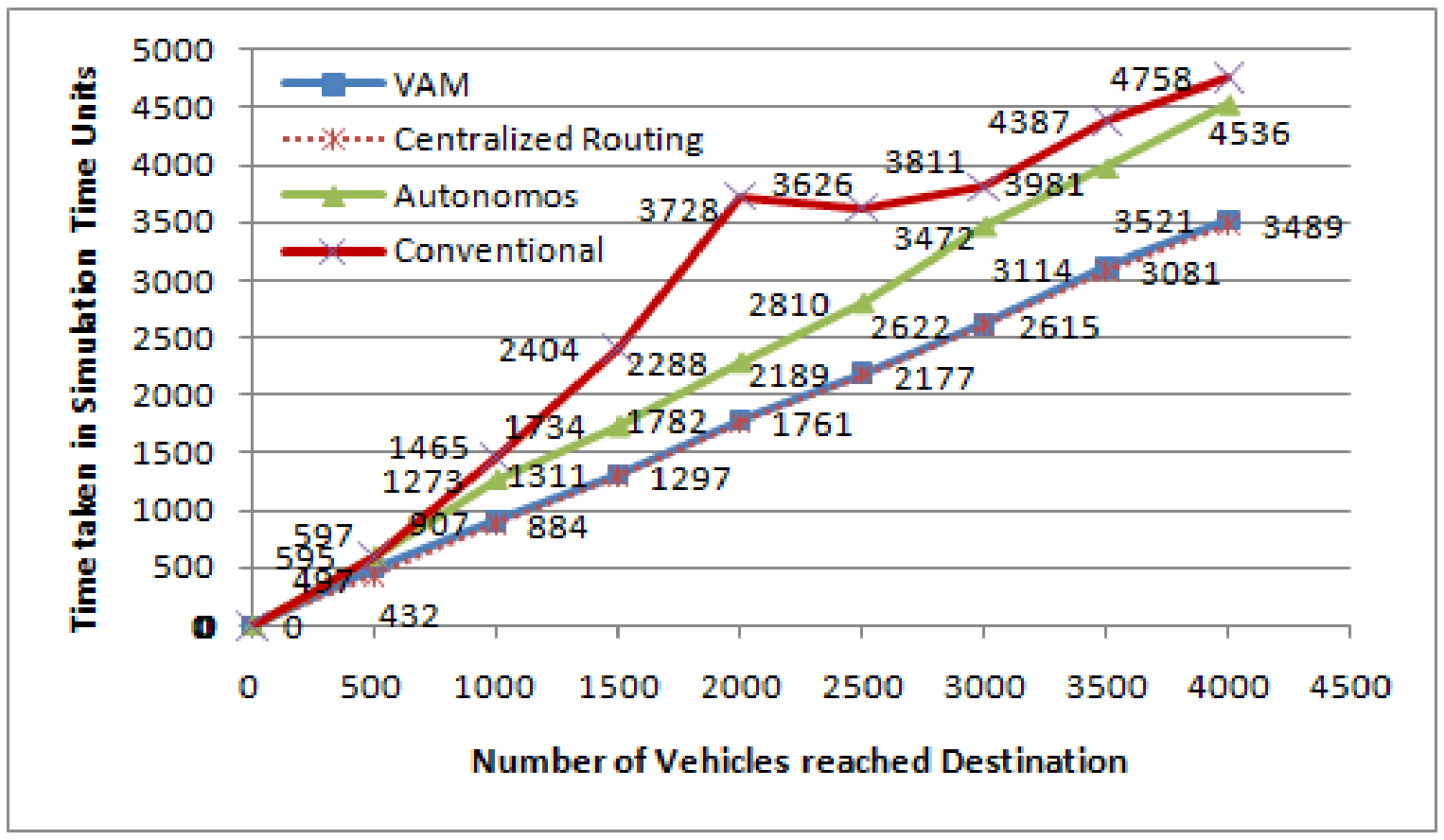

**Figure 5:** Comparing the performance of VAM with contemporary traffic management schemes.

## 5 Conclusion and Open Research Problems

This chapter has focused on recent advances in mitigating traffic congestion and how the development of CAV has introduced a paradigm shift. The notion of dynamic traffic management has been explained and the challenge in applying it has been analyzed. The CAV capabilities are shown to be invaluable and in fact enabler for realizing DTM in practice. Advances made to date have been highlighted and existing techniques have been categorized and compared. We have also presented a new vision, namely, *iRoad*, for applying DTM for scenarios involving a mix of autonomous and human-driven vehicles. Sample preliminary results have discussed as well. The following points out some open issues that are worth further investigation:

- Supporting pedestrians crossing: As pointed out in Section 3, CAV will enable the elimination of traffic signals since vehicles can collaboratively determine the crossing order without causing collisions. With the transition to non-signalized intersections, other road users like pedestrians and bicyclists will become unprivileged; unless the intersection is empty these users will not be able to cross safely. According to the 2017 reports of the US National Highway Transportation Safety Authority (NHTSA) [198], about 6 thousand pedestrians lost their lives due to road crash with intersection crossing being the most frequent cause. Among the road related fatalities in urban areas during 2017, pedestrians represent about 25%. Similarly, about 800 pedal cyclists died in 2017 due to road crashes. Relying on obstacle detection capabilities for autonomous vehicles would not be sufficient for ensuring safety and effective for sustaining traffic flow since, if unregulated, pedestrians crossing could be disorganized and sporadic. In fact, in crowded areas, e.g., downtown Manhattan in New York City, pedestrian crossing at unsignalized intersection could bring traffic to stand still given the number of locals and visitors typically found on the sidewalks. Thus, developing integrated solutions for road crossing is needed, especially in urban setups.
- Integrated DTM Optimization Framework: Road reconfiguration can be a very powerful venue for supporting DTM; as discussed earlier in the chapter, it constitutes a means for closing the control loop. CAV enables a wealth of options such as collaborative intersection crossing and dynamic lane reversal; yet in most publications these options are being exploited individually rather than in an integrated manner. It is important to study which options leverage and conflict with one another and which options are independent. For example, intersection crossing for CAV is more versatile with additional degree of freedom with more reliable and dynamic road data updates and more agile and accurate reaction of vehicles compared to human-driven vehicles. Meanwhile, DLG strategies for CAV provide great flexibility in handling dynamically fluctuated traffic demand. However, applying DLG at non-signalized intersections is still an unresolved issue, and dynamic motion planning algorithms are needed to deal with the phase sequence constraint [105]. A comprehensive optimization framework is needed.
- Handling of Mixed Traffic: CAV enables effective realization of DTM both by providing data and responding to control commands, e.g., to take a system-optimal route. As shown in Section 3 CAV makes it easy for exploiting road reconfiguration as a means for mitigating congestion and dealing with traffic incidents. Nonetheless, human-driven vehicles are still predominant and are expected to coexist on the road with CAV for a long while. Thus, practically CAV-based DTM optimization will have to adapt and prevent the presence of human-driven vehicles from hindering the advantages of CAV. Investigating techniques for DTM in mixed traffic scenarios is necessary. The challenges to be tackled include the fidelity of situation assessment given the diversity of collected data, the effect of driver behavior on trajectory planning for CAV, and scheduling of safe intersection crossing. We envision machine learning techniques to be invaluable in addressing these challenges where CAV are to self-adapt its operation in various settings, e.g., urban or rural, and varying densities of human-driven vehicles.

- Vehicle and Infrastructure Security: In essence DTM is viable only when accurate data are available. Collecting such data can be via numerous means as discussed in Section 2.2. Some of these means such as electromagnetic loops, laser sensors, pressure hoses, and radars, do not identify the vehicles and do not raise privacy concerns. In addition, manipulating these sensors through cyberattacks is not easy giving the physical wiring which mandate the adversary to be present to tamper with sensors. Yet, deploying of these types of sensing is expensive and the trend is to rely on participatory sensing where the data is collected through V2V or V2I communication. However, such data collection methodology raises privacy and security concerns. Basically the location and travel route of the participating vehicles should be shared in order for the system to know demand, detect bottlenecks, and anticipate traffic density. Exposing such information about the individual vehicles constitutes invasion of the drivers/passengers privacy and may be exploited by criminals. For example, knowing that someone left home to a remote destination can be exploited for planning a robbery. Moreover, since the DTM system will aggregate the data collected about individual vehicles, Sybil attacks can be launched by a vehicle to gain on-road privilege. A rogue vehicle could claim multiple identities to inflate the density on a certain road and consequently trigger road reconfiguration that better serves its travel path. For example, by sending messages with different identities to a traffic signal controller or a road-side unit, a vehicle could give the false impression that many vehicles are crowding and more green time or lane reversal is warranted. Thus, guarding DTM systems against cyberattacks is very critical. Securing these systems will be very challenging given the resource constraints, heterogeneity, and mobility of the vehicles. The complexity even grows for traffic involving both autonomous and human-driven vehicles.
- Interaction with Smart City Applications: Recent technological advances in computation and communication devices have revolutionized how people live and interact. The notion of connected communities and smart cities reflects how societies are being transformed. CAV constitutes an example of what one expects in a smart city. Therefore, the scope of DTM needs to be broadened to cope with emerging applications and services that should be continually accessible and efficiently supported while people are commuting. For example, passengers would expect access to entertainment and social media while riding their vehicles. Thus, route selection and even trajectory planning could be influenced by factors that are not dependent on the road conditions. For example, a vehicle may join a platoon to receive streamed movie or prefers a travel path for which the cell phone coverage is at best to access live cast of a game. Similarly, vehicles may decide to stay next to each other to enable peer-to-peer wireless battery charging. One may even prefer scenic routes close to landmarks or passing areas with holiday decorations. These unconventional factors will indeed affect DTM; new strategies are needed as societies are getting more and more modernized.

**Acknowledgement:** Dr. Sookyoung Lee is supported by the National Research Foundation of Korea (NRF) funded by the Korea government (MSIT) (2018R1D1A1B07043671).

## REFERENCES

[1] http://www.nationwide.com/road-congestion-infographic.jsp

[2] https://corporate.tomtom.com/static-files/69e46ddb-4eac-4014-918f-cd5e7cd0c2c4

[3] https://www.weforum.org/agenda/2019/02/commuters-in-these-cities-spend-more-than-8-days-a-year-stuck-in-traffic/

[4] https://static.tti.tamu.edu/tti.tamu.edu/documents/mobility-report-2019.pdf

[5] "21st Century Operations Using 21st Century Technologies", Office of Transportation Operations, U.S. Department of Transportation Federal Highway Administration, http://ops.fhwa.dot.gov/aboutus/one_pagers/opstory.html

[6] J. Ding, Y. Zhang, L. Li, "Accessibility measure of bus transit networks," *IET Intelligent Transportation Systems*, 2(7), pp. 682-688, Feb. 2018.

[7] A. Nuzzolo, and A. Comi, "Advanced public transport and intelligent transport systems: new modelling challenges," Transportm. A: Transp. Sci., Vol. 12, pp. 674-699, 2016.

[8] Y. Dong, S. Wang, L. Li, Z. Zhang, "An empirical study on travel patterns of internet based ride-sharing," *Transport. Res. Part C: Emerg. Technol.,* Vol. 86, pp. 1-22, 2018.

[9] Y.M. Nie, "How can the taxi industry survive the tide of ride-sourcing? Evidence from Shenzhen, China," *Transport. Res. Part C: Emerg. Technol.*, Vol. 79, pp. 242-256, 2017.

[10] https://ops.fhwa.dot.gov/publications/fhwahop13003/index.htm

[11] T. Zeng, O. Semiari, W. Saad, and M. Bennis, "Joint communication and control for wireless autonomous vehicular platoon systems," *IEEE Transactions on Communications*, 2019.

[12] O. Semiari, W. Saad, M. Bennis, and M. Debbah, "Integrated millimeter wave and sub-6 GHz wireless networks: A roadmap for joint mobile broadband and ultra-reliable low-latency communications," *IEEE Wireless Communications*, pp. 1–7, 2019.

[13] D. J. Fagnant and K. Kockelman, "Preparing a nation for autonomous vehicles: opportunities, barriers and policy recommendations," *Transportation Research Part A: Policy and Practice*, vol. 77, pp. 167–181, Jul. 2015.

[14] W. Bernhart and M. Winterhoff, "Autonomous Driving: Disruptive Innovation that Promises to Change the Automotive Industry as We Know It," *Energy Consumption and Autonomous Driving, Proceeding of the 3rd CESA Automotive Electronics Congress*, Springer, 2016.

[15] S. Trommer, E. Fraedrich, V. Kolarova and B. Lenz, "Exploring user expectations on autonomous driving," in the Proceedings of the Automated Vehicles Symposium, San Francisco, USA, June 2016.

[16] T. Litman, "Autonomous vehicles implementation predictions and implications for transport planning," 2018.

[17] S. A. Bagloee, et al. "Autonomous vehicles: challenges, opportunities, and future implications for transportation policies," *Journal of modern transportation*, Vol.24, No.4, pp.284-303, 2016.

[18] R. Hussain, and S. Zeadally. "Autonomous Cars: Research Results, Issues, and Future Challenges," *IEEE Communications Surveys & Tutorials,* Vol.21, No.2, pp.1275-1313, 2018.

[19] https://new.siemens.com/global/en/products/mobility/road-solutions/traffic-management/on-the-road/smart-detection/video-detection.html

[20] P.F. Alcantarilla, M.A. Sotelo, and L. M. Bergasa, "Automatic daytime road traffic control and monitoring system," in the *Proceedings of the 11th IEEE Interational Conf on Intelligent Transportation Systems*, Bejing, China, October 2008.

[21] https://www.lasertech.com/IS-Measure-Traffic-Count.aspx

[22] https://trafficbot.rhythmtraffic.com/in-sync/

[23] https://www.sensourceinc.com/hardware/vehicle-counting-products/

[24] E. Massaro *et al.*, "The Car as an Ambient Sensing Platform [Point of View]," *Proceedings of the IEEE*, vol. 105, no. 1, pp. 3-7, Jan. 2017.

[25] X. Zhang, J. Hong, S. F. Z. Wei, J. Cao, and Y. Ren, "A novel real-time traffic information system based on wireless mesh networks," in the *Proceedings of the IEEE Intelligent Transportation Systems Conference*, Seattle, WA, Sept 2007.

[26] F. Calabrese, M. Colonna, P. Lovisolo, D. Parata, and C. Ratti, "Real-time urban monitoring using cell phones: A case study in Rome," *IEEE Transactions on Intelligent Transportation Systems*, Vol. 12, No. 1, pp.141–151, 2011

[27] https://www.waze.com/

[28] http://inrix.com/

[29] http://www.cellint.com/

[30] Q. Ou, R.L. Bertini, J.W.C. Van Lint, S.P. Hoogendoorn, "A theoretical framework for traffic speed estimation by fusing low-resolution probe vehicle data," *IEEE Transactions on Intelligent Transport. Systems*, Vol. 12, No. 3, pp. 747-756, 2011.

[31] Y. Cheng, X. Qin, J. Jin, B. Ran, "An exploratory shockwave approach to estimating queue length using probe trajectories," *Journal Intelligent Transportation Systems*, Vol. 16, No. 1, pp. 12-23, 2011.

[32] S. Gupte and M. Younis, "Vehicular Networking for Intelligent and Autonomous Traffic Management," in the *Proceedings of the IEEE International Conference on Communications (ICC'12)*, Ottawa, Canada, June 2012.

[33] T. Nadeem, S. Dashtinezhad, C. Liao, and L. Iftode, "Trafficview: a scalable traffic monitoring system," in the *Proceedings of the IEEE International Conference on Mobile Data Management*, pp. 13–26, 2004.
[34] M. D. A. Florides, T. Nadeem, and L. Iftode, "Location-aware services over vehicular ad-hoc networks using car-to-car communication," IEEE Journal on Selected Areas in Communications, Vol. 25, pp. 1590–1602, October 2007.
[35] L. Wischoff, A. Ebner, H. Rohling, M. Lott, and R. Halfmann, "Sotis- a self-organizing traffic information system," in the *Proceedings of the 57th IEEE Semiannual Vehicular Technology Conference (VTC-Spring)*, Vol. 4, pp. 2442–2446, April 2003.
[36] M. Milojevic and V. Rakocevic, "Distributed road traffic congestion quantification using cooperative vanets," in the *Proceedings of 13th Annual Mediterranean in Ad Hoc Networking Workshop (MED-HOC-NET)*, pp. 203–210, June 2014.
[37] M. F. Fahmy and D. N. Ranasinghe, "Discovering dynamic vehicular congestion using VANETs," in the *Proceedings of the 4th International Conference on Information and Automation for Sustainability (ICIAFS 2008)*, Colombo, Sri Lanka, December 2008.
[38] L. Wischhof, A. Ebner, and H. Rohling, "Information dissemination in self-organizing inter-vehicle networks," *IEEE Transactions on Intelligent Transportation Systems*, Vol. 6, No. 1, pp. 90–101, 2005.
[39] R. Bauza, J. Gozalvez, and J. Sanchez-Soriano, "Road traffic congestion detection through cooperative vehicle-to-vehicle communications," in the *Proceedings of the 35th IEEE Conference on Local Computer Networks (LCN 2010)*, pp. 606–612, 2010.
[40] F. Terroso-Sáenz, M. Valdés-Vela, C. Sotomayor-Martínez, R. Toledo-Moreo, and A. F. Gómez-Skarmeta, "A cooperative approach to traffic congestion detection with complex event processing and vanet," *IEEE Transactions on Intelligent Transportation Systems*, Vol. 13, No. 2, pp. 914–929, 2012.
[41] G. Marfia and M. Roccetti, "Vehicular congestion detection and shortterm forecasting: A new model with results," *IEEE Transactions on Vehicular Technology*, Vol. 60, No. 7, pp. 2936–2948, September 2011.
[42] I. Leontiadis, G. Marfia, D. Mack, G. Pau, C. Mascolo, and M. Gerla, "On the effectiveness of an opportunistic traffic management system for vehicular networks," *IEEE Transactions on Intelligent Transportation Systems*, Vol. 12, pp. 1537–1548, 2011.
[43] H.R. Varia, and S.L. Dhingra, "Dynamic optimal traffic assignment and signal time optimization using genetic algorithms," Computer - Aided Civil and Infrastructure Engineering, Vo. 19, No 4, pp. 260–273, 2004.
[44] K. Collins and G. Muntean, "Route-based vehicular traffic management for wireless access in vehicular environments," in the *Proceedings of the 68th IEEE Vehicular Technology Conference (VTC-Spring)*, Calgary, BC, Canada, September 2008.
[45] G. Araujo, F. De L P Duarte-Figueiredo, A. Tostes, and A. Loureiro, "A protocol for identification and minimization of traffic congestion invehicular networks," in the *Proceedings of the Brazilian Symposium on Computer Networks and Distributed Systems (SBRC)*, pp. 103–112, May 2014.
[46] S. C. Nanayakkara, et al., "Genetic algorithm based route planner for large urban street networks," in the Proceedings of the IEEE Congress on Evolutionary Computation (CEC 2007), Singapore, september 2007.
[47] A. Ghazy and T. Ozkul, "Design and simulation of an artificially intelligent vanet for solving traffic congestion," in the Proceedings of the IEEE 6th International Symposium on Mechatronics and its Applications (ISMA '09), Sharjah, United Arab Emirates, March 2009.
[48] Y. Ando, O. Masutani, and S. Honiden., "Performance of pheromone model for predicting traffic congestion," in the Proceedings of the 5th International Joint Conference on Autonomous Agents and Multi-Agent Systems (AAMAS '06), Future University-Hakodate, Japan, May 2006.
[49] A. Ramazani and H. Vahdat-Nejad, "A new context-aware approach to traffic congestion estimation," in the *Proceedings of the 4th International eConference on Computer and Knowledge Engineering (ICCKE)*, pp.504–508, Oct 2014.
[50] T. Ho, and T.H. Heung, "Hierarchical fuzzy logic traffic control at a road junction using genetic algorithms," in the *Proceedings of the IEEE World Congress on Computational Intelligence*, Anchorage, AK, November 1998.
[51] E. Horvitz, J. Apacible, R. Sarin, and L. Liao, "Prediction, expectation, and surprise: Methods, designs, and study of a deployed traffic forecasting service," in the *Proceedings of the 21st Conference on Uncertainty in Artificial Intelligence(UAI 2005)*, Edinburgh, Scotland, July 2005.
[52] T. Hunter, R. Herring, P. Abbeel, and A. Bayen, "Path and travel time inference from GPS probe vehicle data," in the *Proceedings of the NIPS Workshop on Analyzing Networks and Learning with Graphs*, Whistler, BC, Canada, Dec 2009.
[53] D.B. Work, O.P. Tossavainen, S. Blandin, A.M. Bayen, T. Iwuchukwu, and K. Tracton, "An ensemble kalman filtering approach to highway traffic estimation using GPS enabled mobile devices," in the *Proceedings of the 47th IEEE Conference on Decision and Control (CDC 2008)*, Cancun, Mexico, Dec 2008.
[54] A. Bhaskar, T. Tsubota, and E. Chung, "Urban traffic state estimation: Fusing point and zone based data," *Transport. Res. Part C: Emerg. Technol.*, Vol. 48, pp. 120-142, 2014.
[55] M. Zangui, Y. Yin and S. Lawphongpanich, "Differentiated congestion pricing of urban transportation networks with vehicle-tracking technologies," *Transportation Research Part C*, Vol. 36, pp. 434–445, November 2013.

[56] F. Soylemezgiller, M. Kuscu, and D. Kilinc, "A traffic congestion avoidance algorithm with dynamic road pricing for smart cities," in the *Proceedings of the 24th IEEE International Symposium on Personal Indoor and Mobile Radio Communications (PIMRC 2013)*, London, UK, September 2013.
[57] M. Zangui, Y. Yin, and S. Lawphongpanich, "Sensor location problems in path-differentiated congestion pricing," *Transportation Research Part C: Emerging Technologies*, Vol. 55, pp. 217–230, June 2015.
[58] B. Zhou, M. Bliemer, H. Yang and J. He, "A trial-and-error congestion pricing scheme for networks with elastic demand and link capacity constraints," *Transportation Research Part B: Methodological*, Vol. 72, pp. 77-92, February 2015.
[59] J. A. Laval, H. W. Cho, J. C. Muñoz and Y. Yin, "Real-time congestion pricing strategies for toll facilities," *Transportation Research Part B: Methodological*, Vol. 71, pp. 19–31, January 2015.
[60] L. Elefteriadou, S. Washburn, Y. Yin, V. Modi and C. Letter, "Variable Speed Limit (VSL) – Best Management Practice," Final Report to Florida Department of Transportation, 2012.
[61] W. Liu, Y. Yin, and H. Yang, "Effectiveness of variable speed limits considering commuters' long-term response," *Transportation Research Part B: Methodological*, Vol. 81, Part 2, pp. 498–519, November 2015.
[62] Z. Song, Y. Yin, and S. Lawphongpanich, "Optimal deployment of managed lanes in general networks," *International Journal of Sustainable Transportation*, Vol. 9, No. 6, pp. 431–441, 2015.
[63] F. Ahmad, S. Mahmud, G. Khan, and F. Yousaf, "Shortest remaining processing time based schedulers for reduction of traffic congestion," in the *Proceedings of the International Conference on Connected Vehicles and Expo (ICCVE)*, Las Vegas, Nevada, December 2013.
[64] S. Kwatirayo, J. Almhana, and Z. Liu, "Adaptive traffic light control using VANET: A case study," in the *Proceedings of the 9th International Wireless Communications and Mobile Computing Conference (IWCMC 2013)*, pp. 752–757, July 2013.
[65] K. Pandit, D. Ghosal, H. Zhang, and C.-N. Chuah, "Adaptive traffic signal control with vehicular ad hoc networks," *IEEE Transactions on Vehicular Technology*, Vol. 62, No. 4, pp. 1459–1471, May 2013.
[66] K. Al-Khateeb, and J. Johari, "Intelligent dynamic traffic light sequence using RFID," Journal of Computer Science, Vol. 4, No 7, pp. 517–524, 2008.
[67] C. Hu and Y. Wang, "A novel intelligent traffic light control scheme," in the *Proceedings of 9th International Conference on Grid and Cooperative Computing (GCC),* pp. 372-376, 2010.
[68] C. Li and S. Shimamoto, "An Open Traffic Light Control Model for Reducing Vehicles' CO2 Emissions Based on ETC Vehicles," *IEEE Transactions on Vehicular Technology*, Vol. 61, pp. 97–110, 2012.
[69] S. Tomforde, et al., "Decentralised Progressive Signal Systems for Organic Traffic Control," in the *Proceedings of the* 2nd *IEEE International Conference on Self-Adaptive and Self-Organizing Systems (SASO'08)*, Venice, Italy, Oct 2008.
[70] V. Hirankitti, J. Krohkaew, and C. Hogger, "A multi-agent approach for intelligent traffic-light control," *in Proc. of the World Congress on Engineering*, vol. 1., Morgan Kaufmann, London, U.K., 2007.
[71] X. Zheng and L. Chu, "Optimal Parameter Settings for Adaptive Traffic-Actuated Signal Control," in the *Proceedings of 11th International IEEE Conference on Intelligent Transportation Systems (ITSC),* pp. 105-110, 2008.
[72] B. Zhou, J. Cao, and H. Wu, "Adaptive traffic light control of multiple intersections in wsn-based ITS," in the *Proceedings of the 73rd IEEE Vehicular Technology Conference (VTC Spring)*, Yokohama, Japan, May 2011.
[73] S. Azimi, G. Bhatia, R. Rajkumar, and P. Mudalige, "Reliable intersection protocols using vehicular networks," in the *Proceedings of the ACM/IEEE International Conference on Cyber-Physical Systems (ICCPS 2013)*, Philadelphia, PA, April 2013.
[74] Hult, R., Campos, G., Falcone, P., and Wymeersch, H., "An approximate solution to the optimal coordination problem for autonomous vehicles at intersections," in the *Proceedings of American Control Conference*, pp. 763–768. Chicago, IL, USA, 2015.
[75] R. Azimi, G. Bhatia, R. R. Rajkumar, and P. Mudalige, "Stip: Spatio-temporal intersection protocols for autonomous vehicles," in the *Proceedings of the ACM/IEEE International Conference on Cyber-Physical Systems (ICCPS 2014)*, Berlin, Germany, April 2014.
[76] https://www.nsf.gov/awardsearch/showAward?AWD_ID=1446813&HistoricalAwards=false
[77] S. D. Assimonis, T. Samaras and V. Fusco, "Analysis of the microstrip-grid array antenna and proposal of a new high-gain, low-complexity and planar long-range WiFi antenna," *IET Microwaves, Antennas & Propagation*, Vol. 12, No. 3, pp. 332-338, 2018.
[78] "About DD-WRT" www.dd-wrt.com. Retrieved October 25th, 2019.
[79] A. Asadi, Q. Wang, and V. Mancuso. "A survey on device-to-device communication in cellular networks." *IEEE Communications Surveys & Tutorials*, Vol. 16, No. 4, pp. 1801-1819, 2014.
[80] S. Eichler, "Performance evaluation of the IEEE 802.11 p WAVE communication standard," in the *Proceedings of the 66th IEEE Vehicular Technology Conference (VTC-2007 Fall)*, Baltimore, MD, October 2007

[81] A. M. S. Abdelgader, and L. Wu, “The physical layer of the IEEE 802.11 p WAVE communication standard: the specifications and challenges,” in the Proceedings of the World Congress on Engineering and Computer Science (WCECS 2014), Vol. 2. San Francisco, CA, October 2014.
[82] J. B. Kenney “Dedicated short-range communications (DSRC) standards in the United States,” *Proceedings of the IEEE*, Vol. 99, No. 7, pp. 1162-1182, 2011.
[83] A. A. Shahin, and M. Younis, “Alert dissemination protocol using service discovery in Wi-Fi Direct,” in the *Proceedings of the IEEE International Conference on Communications (ICC 2015)*, London, UK, June 2015.
[84] A. Shahin and M. Younis, “Efficient Multi-Group Formation and Communication Protocol for Wi-Fi Direct,” in the *Proceedings of the 40th Annual IEEE Conference on Local Computer Networks (LCN 2015)*, Clearwater Beach, FL, October 2015.
[85] C. Wang, Y.-Q. Tang, “The discussion of system optimism and user equilibrium in traffic assignment with the perspective of game theory,” *Transportation Research Procedia*, Vol. 25, pp. 2970-2979, 2017.
[86] F. Kessels, *Traffic Flow Modelling: Introduction to Traffic Flow Theory Through a Genealogy of Models*, Springer, 2019.
[87] https://ops.fhwa.dot.gov/trafficanalysistools/type_tools.htm
[88] Hao Xu, Hongchao Liu, Huaxin Gong, “Modeling the asymmetry in traffic flow (a): Microscopic approach,” Applied Mathematical Modelling, Vol. 37, No. 22, pp. 9431-9440, 2013.
[89] Rosa M Velasco and Patricia Saavedra, “Macroscopic Models in Traffic Flow,” *Qualitative Theory of Dynamical Systems*, Vol. 7, No. 1, pp. 237-252, July 2008.
[90] R. Aghamohammadi, J. A. Laval, “Dynamic traffic assignment using the macroscopic fundamental diagram: A Review of vehicular and pedestrian flow models,” *Transportation Research Part B: Methodological*, https://doi.org/10.1016/j.trb.2018.10.017, 2018 (to appear).
[91] A. Spiliopoulou, M. Kontorinaki, M. Papageorgiou, P. Kopelias, “Macroscopic traffic flow model validation at congested freeway off-ramp areas,” *Transportation Research Part C: Emerging Technologies*, Vol. 41, pp. 18-29, 2014.
[92] G. Costeseque, and A. Duret, “Mesoscopic multiclass traffic flow modeling on multi-lane sections,” in the *Proceedings of Transportation Research Board 95th Annual Meeting*, Washington DC, January 2016.
[93] B. N. Janson, “Dynamic traffic assignment for urban road networks,” *Transportation Research Part B: Methodological*, Vol. 25, No. 2–3, pp. 143-161, 1991.
[94] Y. Wang, W.Y. Szeto, K. Han, and T. L. Friesz, “Dynamic traffic assignment: A review of the methodological advances for environmentally sustainable road transportation applications,” *Transportation Research Part B: Methodological*, Vol. 111, pp. 370-394, 2018.
[95] A. Hall, S. Hippler, and M. Skutella, “Multicommodity flows over time: Efficient algorithms and complexity,” *Theoretical Computer Science*, Vol. 379, No. 3, pp. 387-404, 2007.
[96] W. L. Gisler, and N. J. Rowan, “Development of Fiberoptic Sign Displays for Dynamic Lane Assignment,” *Technical report*, Texas Transportation Institute, Austin, Texas, Jun. 1992.
[97] L. Zhang, and G. Wu, "Dynamic lane grouping at isolated intersections: problem formulation and performance analysis," Transportation Research Record, Vol.2311, No.1 pp.152-166, 2012.
[98] W. KM Alhajyaseen, et al., "The effectiveness of applying dynamic lane assignment at all approaches of signalized intersection," Case studies on transport policy Vol.5, Issue 2, pp.224-232, 2017.
[99] K. J. Assi, and N. T. Ratrout. "Proposed quick method for applying dynamic lane assignment at signalized intersections," IATSS research, Vol.42, No.1, pp.1-7, 2018.
[100] W. KM Alhajyaseen, et al., "The integration of dynamic lane grouping technique and signal timing optimization for improving the mobility of isolated intersections," Arabian Journal for Science and Engineering, Vol.42, NO.3, pp.1013-1024, 2017.
[101] X. Li, H. Wang, and J. Chen, "Dynamic lane-use assignment model at signalized intersections under tidal flow," ICTE 2013: Safety, Speediness, Intelligence, Low-Carbon, Innovation, pp.2673-2678, 2013.
[102] Z. Zhong, et al., "An optimization method of dynamic lane assignment at signalized intersection," IEEE International Conference on Intelligent Computation Technology and Automation (ICICTA), Vol. 1, 2008.
[103] Wu, Guoyuan, et al. "Simulation-based benefit evaluation of dynamic lane grouping strategies at isolated intersections," 2012 15th International IEEE Conference on Intelligent Transportation Systems, 2012.
[104] X. Jiang, R. Jagannathan, and D. Hale, "Dynamic Lane Grouping at Signalized Intersections: Selecting the Candidates and Evaluating Performance," Institute of Transportation Engineers, ITE Journal, Vol. 85, No.11, 2015.
[105] W. Weili, J. Zheng, and H. X. Liu., "A capacity maximization scheme for intersection management with automated vehicles," Transportation research procedia, Vol.23, pp.121-136, 2017.
[106] S. Kim, S. Shekhar, and M. Min, "Contraflow transportation network reconfiguration for evacuation route planning," *IEEE Transactions on Knowledge and Data Engineering* Vol.20, No.8, pp.1115-1129, 2008.

[107] Xie, Chi, and Mark A. Turnquist, "Lane-based evacuation network optimization: An integrated Lagrangian relaxation and tabu search approach." *Transportation Research Part C: Emerging Technologies*, Vol.19, No.1, pp.40-63, 2011.

[108] Zhang, Xin Ming, Shi An, and Bing Lei Xie, "A cell-based regional evacuation model with contra-flow lane deployment," *Advanced Engineering Forum*, Vol.5, *Trans Tech Publications*, 2012.

[109] J.W. Wang, et al., "Evacuation planning based on the contraflow technique with consideration of evacuation priorities and traffic setup time." *IEEE transactions on intelligent transportation systems*, Vol.14, No.1, pp.480-485, 2012.

[110] J. Hua, et al. "An integrated contraflow strategy for multimodal evacuation," *Mathematical Problems in Engineering*, Vol.2014, 2014.

[111] U. Pyakurel, and Stephan Dempe. "Earliest Arrival Flow with Partial Lane Reversals for Evacuation Planning," *International Journal of Operational Research/Nepal (IJORN)*, Vol.8, No.1, 2019.

[112] Dhamala, Tanka Nath, Urmila Pyakurel, and Ram Chandra Dhungana. "Abstract contraflow models and solution procedures for evacuation planning," *Journal of Mathematics Research*, Vol.10, No.4, pp.89-100, 2018.

[113] U. Pyakurel, S. Dempe, and T.N. Dhamala, "Efficient algorithms for flow over time evacuation planning problems with lane reversal strategy," *TU Bergakademie Freiberg*., 2018

[114] U. Pyakurel, H.N. Nath, and T.N. Dhamala, "Partial Contraflow with Path Reversals for Evacuation Planning," *Annals of Operations Research*, 2018.

[115] W.W. Zhou, et al., "An intelligent traffic responsive contraflow lane control system," *Proceedings of VNIS'93-Vehicle Navigation and Information Systems Conference*, 1993.

[116] M. Hausknecht, et al., "Dynamic lane reversal in traffic management," *2011 14th International IEEE Conference on Intelligent Transportation Systems (ITSC)*, 2011.

[117] T. Lu, Z. Yang, D. Ma, and S. Jin, "Bi-Level Programming Model for Dynamic Reversible Lane Assignment," *IEEE Access*, Vol.IL6, pp.71592-71601, 2018.

[118] M. Hausknecht, T. Au, and P. Stone. "Autonomous intersection management: Multi-intersection optimization," *2011 IEEE/RSJ International Conference on Intelligent Robots and Systems(IROS)*, San Francisco, USA, September 2011.

[119] M. Duell, et al. "System optimal dynamic lane reversal for autonomous vehicles," *2015 IEEE 18th International Conference on Intelligent Transportation Systems*, 2015.

[120] M. W. Levin, and Stephen D. Boyles, "A cell transmission model for dynamic lane reversal with autonomous vehicles." *Transportation Research Part C: Emerging Technologies*, Vol. 68, pp. 126-143, 2016.

[121] K.F. Chu, A. Y. S. Lam, and V. OK Li, "Dynamic lane reversal routing and scheduling for connected autonomous vehicles." In the Proceedings of the *International Smart Cities Conference (ISC2)*, Wuxi, China, Sept. 2017.

[122] X. Li, J. Chen, and H. Wang, "Study on Flow Direction Changing Method of Reversible Lanes on Urban Arterial Roadways in China," *Procedia - Social and Behavioral Sciences*, Vol. 96, pp.807-816, 2013.

[123] C. Krause, N. Kronpraset, J. Bared, and W. Zhang, "Operational advantages of dynamic reversible left-lane control of existing signalized diamond interchanges," *Journal of Transportation Engineering*, Vol. 141, No. 5, 2014.

[124] J. Zhao, Yue Liu, and X. Yang, "Operation of signalized diamond interchanges with frontage roads using dynamic reversible lane control," *Transportation research part C: emerging technologies*, Vol. 5, pp. 196-209, 2015.

[125] J. Zhao, et al. "Increasing the capacity of signalized intersections with dynamic use of exit lanes for left-turn traffic," *Journal of the Transportation Research Board*, Vol. 2355, No.1, pp. 49-59, 2013.

[126] H. Mirzaei, and T. Givargis. "Fine-grained acceleration control for autonomous intersection management using deep reinforcement learning," In the Proceedings of *IEEE SmartWorld, Ubiquitous Intelligence & Computing, Advanced & Trusted Computed, Scalable Computing & Communications, Cloud & Big Data Computing, Internet of People and Smart City Innovation (SmartWorld/SCALCOM/UIC/ATC/CBDCom/IOP/SCI)*, San Francisco, CA, Aug. 2017.

[127] J. Wu, A. Abbas-Turki, and A. El Moudni, "Cooperative driving: an ant colony system for autonomous intersection management." *Applied Intelligence*, Vol. 37, No. 2, pp. 207-222, 2012.

[128] D. Carlino, S. D. Boyles, and P. Stone, "Auction-based autonomous intersection management." In the Proceedings of the *16th International IEEE Conference on Intelligent Transportation Systems (ITSC 2013)*, 2013.

[129] G. R. de Campos, et al., "Traffic coordination at road intersections: Autonomous decision-making algorithms using model-based heuristics," *IEEE Intelligent Transportation Systems Magazine*, Vol. 9, No. 1, pp. 8 – 21, 2017.

[130] A. A. Malikopoulos, C. G. Cassandras, and Y. J. Zhang, "A decentralized energy-optimal control framework for connected automated vehicles at signal-free intersections," *Automatica*, Vol. 93, pp. 244-256, 2018.

[131] F. Altché, X. Qian, and A. de La Fortelle, "An algorithm for supervised driving of cooperative semi-autonomous vehicles (extended)." arXiv preprint arXiv:1706.08046, 2017.

[132] T.-C. Au, S. Zhang, and P. Stone, "Autonomous intersection management for semi-autonomous vehicles." *Routledge Handbook of Transportation*. Routledge, pp. 116-132, 2015.

[133] J. Lee, and B. Park, "Development and evaluation of a cooperative vehicle intersection control algorithm under the connected vehicles environment." *IEEE Transactions on Intelligent Transportation Systems*, Vol. 13, No. 1, pp. 81-90, 2012.

[134] Q. Lu, and K.-D. Kim, "Intelligent intersection management of autonomous traffic using discrete-time occupancies trajectory." *Journal of Traffic and Logistics Engineering*, Vol 4.1 (2016): 1-6.
[135] J. H. Dahlberg, and V. Tuul, "Intelligent Traffic Intersection Management Using Motion Planning for Autonomous Vehicles", 2017.
[136] M. Pourmehrab, et al., "Optimizing signalized intersections performance under conventional and automated vehicles traffic." *IEEE Transactions on Intelligent Transportation Systems*, 2019.
[137] Q, Lu, and K.-D. Kim, "Autonomous and connected intersection crossing traffic management using discrete-time occupancies trajectory." *Applied Intelligence*, Vol. 49, No.5, pp. 1621-1635, 2019.
[138] X. Meng, and C. G. Cassandras, "A Real-Time Optimal Eco-driving for Autonomous Vehicles Crossing Multiple Signalized Intersections." *arXiv preprint arXiv:1901.11423,* 2019.
[139] P. Li, and X. Zhou, "Recasting and optimizing intersection automation as a connected-and-automated-vehicle (CAV) scheduling problem: A sequential branch-and-bound search approach in phase-time-traffic hypernetwork." *Transportation Research Part B: Methodological*, Vol. 105, pp. 479-506, 2017.
[140] G. Sharon, S. D. Boyles, and P. Stone, "Intersection management protocol for mixed autonomous and human-operated vehicles." *Transportation Research Part C: Emerging Technologies*, 2017.
[141] C. Wuthishuwong, and A. P. Traechtler, "Consensus-based local information coordination for the networked control of the autonomous intersection management." Complex & Intelligent Systems 3.1 (2017): 17-32.
[142] J. J. B. Vial, et al. "Scheduling autonomous vehicle platoons through an unregulated intersection." *arXiv preprint arXiv:1609.04512*, 2016.
[143] P. Dai, et al. "Quality-of-experience-oriented autonomous intersection control in vehicular networks." *IEEE Transactions on Intelligent Transportation Systems*, Vol. 17, No. 7, pp. 1956-1967, 2016.
[144] Q. Guo, L. Li, and X. Ban, "Urban traffic signal control with connected and automated vehicles: A survey," *Transportation Research Part C: Emerging Technologies*, Vol. 101, pp. 313-334, 2019.
[145] C. Hu and Y. Wang, "A novel intelligent traffic light control scheme," in the *Proceedings of the 9th International Conference on Grid and Cooperative Computing (GCC)*, pp. 372-376, 2010.
[146] B. Zhou, J. Cao, X. Zeng, and H. Wu, "Adaptive traffic light control in wireless sensor network-based intelligent transportation system," in the *Proceedings of the 72nd IEEE Vehicular Technology Conference*, Ottawa, Canada, 2010.
[147] S. Sameh, A. El-Mahdy, and Y. Wada, "A Linear Time and Space Algorithm for Optimal Traffic-Signal Duration at an Intersection," *IEEE Transactions on Intelligent Transportation Systems* 16.1, 387-395, 2015.
[148] H. Prothmann, et al., "Organic traffic light control for urban road networks," *International Journal of Autonomous and Adaptive Communications Systems*, Vol. 2, pp. 203-225, 2009.
[149] S. Tomforde, H. Prothmann, J. Branke, J. H¨ahner, C. M¨uller-Schloer, and H. Schmeck, "Possibilities and limitations of decentralised traffic control systems," Proc. of the IEEE World Congress on Computational Intelligence, pp. 3298–3306, Barcelona, Spain, July 2010.
[150] S. Lee, M. Younis, A. Murali, and M. Lee, "Dynamic Local Vehicular Flow Optimization Using Real-time Traffic Conditions at Multiple Road Intersections," IEEE ACCESS, PP(99):1-1, February 2019.
[151] Mario Collotta, Lucia Lo Bello and Giovanni Pau. "A novel approach for dynamic traffic lights management based on Wireless Sensor Networks and multiple fuzzy logic controllers." *Expert Syst. Appl.*, Vol. 42, pp. 5403-5415, 2015
[152] M. Elgarej, M. Khalifa, and M. Youssfi. "Traffic Lights Optimization with Distributed Ant Colony Optimization Based on Multi-agent System," *International Conference on Networked Systems*. Springer International Publishing, 2016.
[153] Samah El-Tantawy, Baher Abdulhai, and Hossam Abdelgawad, "Multiagent reinforcement learning for integrated network of adaptive traffic signal controllers (marlin-atsc): methodology and large-scale application on downtown Toronto," IEEE Transactions on Intelligent Transportation Systems, 14(3):1140–1150, 2013.
[154] Ivana Dusparic and Vinny Cahill, "Autonomic multi-policy optimization in pervasive systems: Overview and evaluation," ACM Transactions on Autonomous and Adaptive Systems (TAAS), 7(1):11, 2012.
[155] Ana LC Bazzan, Denise de Oliveira, and Bruno C da Silva, "Learning in groups of traffic signals," Engineering Applications of Artificial Intelligence, 23(4):560–568, 2010.
[156] Monireh Abdoos, Nasser Mozayani, and Ana LC Bazzan, "Holonic multiagent system for traffic signals control," Engineering Applications of Artificial Intelligence, 26(5-6):1575–1587, 2013.
[157] Xie, X.-F.; Smith, S. F.; and Barlow, G. J., "Schedule-driven coordination for real-time traffic network control," In 22nd International Conference on Automated Planning and Scheduling (ICAPS), pp. 323–331, 2012.
[158] P. Varaiya, "Smart cars on smart roads. Problems of control," IEEE Transactions on Automatic Control, Vol. 38, No. 2, pp. 195–207, Feb. 1993.
[159] C. Katrakazas, M. Quddus, W.-H. Chen, and L. Deka, "Real-time motion planning methods for autonomous on-road driving: State-of-the-art and future research directions," *Transportation Research Part C: Emerging Technologies*, Vol. 60, pp. 416-442, 2015.

[160] S. Zhang, W. Deng, Q. Zhao, H. Sun and B. Litkouhi, "Dynamic Trajectory Planning for Vehicle Autonomous Driving," in the *Proceedings of the IEEE International Conference on Systems, Man, and Cybernetics*, Manchester, UK, pp. 4161-4166, October 2013

[161] L. Ma, J. Yang, and M. Zhang, "A two-level path planning method for on-road autonomous driving," in the Proceedings of the 2nd International Conference on Intelligent System Design and Engineering Application, pp. 661–664, Sanya, Hainan, China, January 2012.

[162] J. Lee and B. Park, "Development and evaluation of a cooperative vehicle intersection control algorithm under the connected vehicles environment," *IEEE Transactions on Intelligent Transportation Systems*, Vol. 13, No. 1, pp. 81–90, 2012.

[163] Z. Li, et al. "Temporal-Spatial Dimension Extension-Based Intersection Control Formulation for Connected and Autonomous Vehicle Systems", *Transportation Research: Part C Emerging Technologies*, Vol. 104, pp. 234-248, June 2019.

[164] D. Chen, S. Ahn, M. Chitturi, and D. A. Noyce, "Towards vehicle automation: Roadway 32 capacity formulation for traffic mixed with regular and automated vehicles," Transportation research part B: methodological, Vol. 100, pp. 196–221, 2017.

[165] H. Jiang, J. Hu, S. An, M. Wang, and B. B. Park, "Eco approaching at an isolated signalized intersection under partially connected and automated vehicles environment," *Transportation Research Part C: Emerging Technologies*, Vol. 79, pp. 290-307, 2017.

[166] J. Ma, X. Li, F. Zhou, J. Hu, and B. B. Park, "Parsimonious shooting heuristic for trajectory design of connected automated traffic part II: Computational issues and optimization," *Transportation Research Part B: Methodological*, vol. 95, pp. 421–441, Jan. 2017.

[167] M. Wang, T. Ganjineh, R. Rojas, "Action annotated trajectory generation for autonomous maneuvers on structured road networks," In the *Proceedings of the 5th International Conference on Automation, Robotics and Applications*, pp. 67–72, 2011.

[168] T. Gu, and J.M. Dolan, "Toward human-like motion planning in urban environments," in the Proceedings of the IEEE Intelligent Vehicles Symposium (IV 2014), pp. 350–355, Dearborn, MI, June 2014.

[169] Y. Wei, et al., "Dynamic programming-based multi-vehicle longitudinal trajectory optimization with simplified car following models," *Transportation Research Part B: Methodological*, Vol. 106, pp. 102-129, 2017.

[170] F. Zhou, X. Li, and J. Ma, "Parsimonious shooting heuristic for trajectory design of connected automated traffic part I: Theoretical analysis with generalized time geography," *Transportation Research Part B: Methodological*, Vol. 95, pp. 394-420, 2017.

[171] L. Chai, B. Cai, W. ShangGuan, J. Wang, and H. Wang, "Connected and autonomous vehicles coordinating approach at intersection based on space–time slot," *Transportmetrica A: Transport Science*, Vol. 14, No. 10, pp. 929-951, 2018.

[172] K. Dresner and P. Stone, "A multi-agent approach to autonomous intersection management," *J. Artif. Int. Res.,* Vol. 31, No. 1, pp. 591-656, March 2008.

[173] D. Chen, A. Srivastava, S. Ahn, and T. Li, "Traffic dynamics under speed disturbance in mixed traffic with automated and non-automated vehicles," *Transportation Research Part C: Emerging Technologies*, 2019 (https://doi.org/10.1016/j.trc.2019.03.017).

[174] S. Gong, J. Shen, and L. Du, "Constrained optimization and distributed computation based car following control of a connected and autonomous vehicle platoon," *Transportation Research Part B: Methodological*, Vol. 94, pp.314-334, 2016.

[175] W. Gao, Z.P. Jiang, and, K. Ozbay, "Data-driven adaptive optimal control of connected vehicles," *IEEE Transactions on Intelligent Transportation Systems*, Vol. 18, No. 5, pp.1122-1133, 2017.

[176] S. Lefèvre, A. Carvalho, and F. Borrelli, "A learning-based framework for velocity control in autonomous driving. IEEE Transactions on Automation Science and Engineering, Vol. 13, No. 1, pp.32-42, 2016.

[177] A. Talebpour, H. Mahmassani, and S. Hamdar, "Multi-regime sequential risk-taking model of car-following behavior: specification, calibration, and sensitivity analysis," *Transportation Research Record: Journal of the Transportation Research Board*, (2260), Vol 2260, No. 1, pp.60-66, 2011.

[178] T. Kosch, C. Schroth, M. Strassberger, and M. Bechler, *Automotive Inter-networking*, Wiley's, UK, April 9, 2012.

[179] H. Hartenstein, and K. Laberteaux, *VANET Vehicular Applications and Inter-Networking Technologies*, Wiley's, UK, February 2010.

[180] F. D. Da Cunha, A. Boukerche, L. Villas, A. Viana, and A. A. F. Loureiro. "Data Communication in VANETs: A Survey, Challenges and Applications," *Research Report RR-8498*, INRIA Saclay, 2014

[181] T. Kitani, T. Shinkawa, N. Shibata, K. Yasumoto, M. Ito, and T. Higashinoz, "Efficient vanet-based traffic information sharing using buses on regular routes," in the *Proceedings IEEE Vehicular Technology Conference (VTC Spring)*, pp. 3031–3036, May 2008.

[182] J. Pan, I. S. Popa, K. Zeitouni, and C. Borcea, “Proactive Vehicular Traffic Re-routing for Lower Travel Time,” *IEEE Transactions on Vehicular Technology*, Vol. 62, No. 8, pp. 3551–3568, October 2013.

[183] A. Dua, N. Kumar, and S. Bawa, “A systematic review on routing protocols for Vehicular Ad Hoc Networks, Vehicular Communications,” Vol. 1, No. 1, pp. 33-52, January 2014.

[184] F. Ahmad, S. Mahmud, G. Khan, and F. Yousaf, “Shortest remaining processing time based schedulers for reduction of traffic congestion,” in the *Proceedings of the International Conference on Connected Vehicles and Expo (ICCVE)*, Las Vegas, Nevada, December 2013.

[185] S. Kwatirayo, J. Almhana, and Z. Liu, “Adaptive traffic light control using VANET: A case study,” in the *Proceedings of the 9th International Wireless Communications and Mobile Computing Conference (IWCMC 2013)*, pp. 752–757, July 2013.

[186] K. Pandit, D. Ghosal, H. Zhang, and C.-N. Chuah, “Adaptive traffic signal control with vehicular ad hoc networks,” *IEEE Transactions on Vehicular Technology*, Vol. 62, No. 4, pp. 1459–1471, May 2013.

[187] X. Zhang, J. Hong, S. F. Z. Wei, J. Cao, and Y. Ren, “A novel real-time traffic information system based on wireless mesh networks,” in the *Proceedings of the IEEE Intelligent Transportation Systems Conference*, Seattle, WA, Sept 2007.

[188] “Report to Congress on Catastrophic Hurricane Evacuation Plan Evaluation,” U.S. Department of Transportation, June 2006. http://www.fhwa.dot.gov/reports/hurricanevacuation/

[189] A. Shahin, and M. Younis, “A Framework for P2P Networking of Smart Devices Using Wi-Fi Direct,” in the *Proceedings of the 25th IEEE International Symposium on Personal, Indoor and Mobile Radio Communications (PIMRC 2014)*, Washington, DC, September 2014.

[190] A. Shahin and M. Younis, “Alert Dissemination Protocol Using Service Discovery in Wi-Fi Direct,” in the *Proceedings of the IEEE International Conference on Communications (ICC 2015)*, London, UK, June 2015.

[191] A. Shahin and M. Younis, “Efficient Multi-Group Formation and Communication Protocol for Wi-Fi Direct,” in the *Proceedings of the 40th Annual IEEE Conference on Local Computer Networks (LCN 2015)*, Clearwater Beach, FL, October 2015.

[192] K. Rabieh, M. Mahmoud, T. Guo and M. Younis, “Privacy-Preserving Route Reporting Scheme for Traffic Management in VANETs”, in the *Proceedings of the IEEE International Conference on Communications (ICC 2015)*, London, UK, June, 2015.

[193] K. Rabieh, M. Mahmoud, T. Guo and M. Younis, “Cross-Layer Scheme for Detecting Large-scale Colluding Sybil Attack in VANETs”, in the *Proceedings of the IEEE International Conference on Communications (ICC 2015)*, London, UK, June, 2015.

[194] S. Olariu, M. Eltoweissy, and M. Younis, “Towards Autonomous Vehicular Clouds,” ICST Transactions on Mobile Communications and Applications, Vol., 11, No. 7-9, e2, 2011.

[195] J. Douceur, “The Sybil Attack,” in the *Proceedings of the International Workshop on Peer-to-Peer Systems (IPTPS)*, March 2002.

[196] M. Raya and J.-P. Hubaux. “The Security of Vehicular Networks,” in the *Proceedings of the 3rd ACM workshop on Security of ad hoc and sensor networks (SASN 2005)*, pp. 11-21, 2005.

[197] A. Wegener, et al., “Designing a decentralized traffic information system - autonomos,” in the *Proceedings of the 16th ITG/GI - Fachtagung Kommunikation in Verteilten Systemen (KiVS)*, Kassel, Germany, March 2009.

[198] US National Highway Transportation Safety Authority, Traffic Safety Facts 2017, https://crashstats.nhtsa.dot.gov/Api/Public/ViewPublication/812806